\documentclass[12pt,a4paper]{article}

\usepackage{latexsym}
\usepackage[all]{xy}
\usepackage{amsmath}
\usepackage{amsfonts}
\usepackage{amssymb}

\newcommand{\be}{\begin{equation}}
\newcommand{\ee}{\end{equation}}
\newcommand{\bea}{\begin{eqnarray}}
\newcommand{\eea}{\end{eqnarray}}

\newcommand{\qedsymb}{{\em Q.E.D.}}

\renewcommand{\theequation}{\arabic{section}.\arabic{equation}}

\def\cG{{\cal G}}                       %
\def\H{{\cal H}}                        %
\def\cO{{\cal O}}                       %
\def\cH{{\mathcal H}}                   %
\def\cF{{\mathcal F}}                   %
\def\ri{{\mathrm{i}}}                   %
\def\cR{{\cal R}}                       %
\def\bR{{\mathbb R}}                    %
\def\bC{{\mathbb C}}                    %
\def\bZ{{\mathbb Z}}                    %
\def\diag{\mathrm{diag}}                %
\def\red{\mathrm{red}}                  %
\def\bT{{\mathbb T}}                    %
\def\cU{{\cal U}}                       %
\def\cV{{\cal V}}                       %
\def\fC{\mathfrak{C}}                   %
\def\hp{{\hat p}}                       %
\def\cD{{\cal D}}                       %
\def\cZ{{\cal Z}}                       %
\def\cI{{\cal I}}                       %
\def\cA{{\mathcal A}}                   %
\def\cJ{{\mathcal J}}                   %
\def\mI{{\mathrm{I}}}                   %
\def\mII{{\mathrm{II}}}                 %
\def\cP{{\mathcal P}}                   %
\def\cQ{{\mathcal Q}}                   %
\def\1{{\mbox{\boldmath $1$}}}          %
\def\0{{\mbox{\boldmath $0$}}}          %

\begin{document}

\vspace*{0.5cm}
\begin{center}
{\Large \bf Trigonometric Sutherland systems and their Ruijsenaars
duals from symplectic reduction}
\end{center}

\vspace{0.2cm}

\begin{center}
L. Feh\'er${}^{a,b}$ and  V.~Ayadi${}^{b}$
 \\

\bigskip

${}^{a}$Department of Theoretical Physics, MTA  KFKI RMKI\\
H-1525 Budapest, P.O.B. 49,  Hungary \\
e-mail: lfeher@rmki.kfki.hu

\bigskip

${}^b$Department of Theoretical Physics, University of Szeged\\
Tisza Lajos krt 84-86, H-6720 Szeged, Hungary\\
e-mail: ayadi.viktor@stud.u-szeged.hu

\bigskip

\end{center}

\vspace{0.2cm}

\begin{abstract}
Besides its usual interpretation as a system of $n$
indistinguishable particles moving on the circle, the trigonometric
Sutherland system can be viewed alternatively as a system of
distinguishable particles on the circle or on the line, and  these 3
physically distinct systems are in duality with corresponding
variants of the rational Ruijsenaars-Schneider system. We explain
that the 3 duality relations, first obtained by Ruijsenaars in
1995, arise naturally from the Kazhdan-Kostant-Sternberg symplectic
reductions of the cotangent bundles of the group $U(n)$ and its
covering groups $U(1) \times SU(n)$ and $\bR \times SU(n)$,
respectively.  This geometric interpretation enhances our understanding of
the duality relations and simplifies
Ruijsenaars' original direct arguments that led to their discovery.
\end{abstract}

\newpage

\section{Introduction}
\setcounter{equation}{0}

We deal here with certain aspects of integrable classical many-body systems in one spatial dimension, which
is an important area of mathematical physics having many applications and connections to other areas
as reviewed, for example, in \cite{OPI,SR-CRM,Cal,Suth,Eti}.

In the impressive series of papers \cite{SR-CMP,RIMS94,RIMS95} Ruijsenaars established an
intriguing duality relation among
Calogero type integrable many-body systems.
The phase spaces of the dual pairs of systems are related by a symplectomorphism that identifies
the action variables of the `first' system as the  particle-positions of the `second' system,
and vice versa.
The duality map, alias the action-angle transform,  was constructed in \cite{SR-CMP, RIMS94, RIMS95}  by direct means
for each non-elliptic system associated with the $A_n$ root system,
covering both the standard non-relativistic systems and their relativistic deformations \cite{RS}.
It was used to analyze the classical dynamics of the systems, and later it was also shown
to have a quantum mechanical counterpart, the so-called bispectral property \cite{DG, SR-Kup}.

The self-dual character of the rational
Calogero system was already noted by Kazhdan, Kostant and Sternberg (KKS) in their
famous paper \cite{KKS} that introduced the tool of
symplectic reduction into the study of
Calogero type systems.
In fact, Ruijsenaars based his analysis on certain algebraic relations satisfied
by the pertinent Lax matrices which are reminiscent  of moment map constraints, and hinted that
there might lurk a KKS type symplectic reduction picture behind the duality in general.
This conjecture was later vigorously advocated in the work of Gorsky-Nekrasov and their collaborators
\cite{GN,Go,Arut1,Nekr,FR, Fock+,GM}, but
it was not fully substantiated
since the main concern of these authors was the reduction
of infinite-dimensional phase spaces that pose serious technical difficulties.
Their work contains interesting ideas also about finite-dimensional reductions
related to  duality.
The elaboration of these ideas requires further effort since the topological subtleties and
the distinctions between the complex phase spaces and their  different real forms were
swept under the carpet in the, otherwise quite remarkable, articles cited above.

One of us, jointly with C.~Klim\v c\'\i k, recently
explored the finite-dimensional symplectic reductions
that explain the duality between
the hyperbolic Sutherland and the rational Ruijsenaars-Schneider
systems \cite{FK-JPA} as well as the duality between
two real forms of the trigonometric Ruijsenaars-Schneider system \cite{FK3}.
The present paper is devoted to other cases, namely,  to the duality relations involving different
variants of the trigonometric Sutherland system and their Ruijsenaars duals introduced
originally in \cite{RIMS95}.
The ideas that we shall use here are similar to those applied in \cite{FK-JPA, FK3}, and  basic observations
were pointed out previously in \cite{Go,Nekr,Fock+}. However, the details are quite non-trivial
and  each
case needs a separate analysis.
One cannot simply apply  analytic continuation  or degeneration
of one case to the other, since there are considerable topological and analytical subtleties
that cannot be handled by such methods.

It is necessary to recall that the trigonometric Sutherland system,
formally given by the Hamiltonian
\be
H_{\mathrm{Suth}}(q,p)=\frac{1}{2} \sum_{i=1}^n p_i^2 + \frac{1}{4} \sum_{1\leq i<j\leq n}
 \frac{x^2}{\sin^2\!\left(\frac{q_i - q_j}{2}\right)},
\label{1.1}\ee
with real coupling parameter $x$,
admits 3 physically different variants depending on the choice of the domain of the
position variables \cite{RIMS95}. Specifically, one can view it most naturally as a system of $n$ indistinguishable
particles moving on the circle, or as systems of distinguishable particles either on the
circle or one the line.
Respectively, the configuration spaces are chosen from the list
\be
Q(n),
\quad
U(1)\times SQ(n),
\qquad
\bR \times SQ(n),
\label{1.2}\ee
where $Q(n)$ belongs to indistinguishable particles, while in the latter two cases
one can separate the freedom of center of mass motion either as $U(1) \simeq S^1$ or
as $\bR$, and $SQ(n)$ is the arena of the motion relative to the center of mass.
In fact, $Q(n)$ can be realized as $Q(n)\equiv \bT(n)^0/S(n)$, where
$\bT(n)^0$ is the regular part of the standard
maximal torus of $U(n)$ on which the symmetric (Weyl) group $S(n)$ acts, and
$SQ(n)$ is similarly related to the group $SU(n)$.
The third configuration space in (\ref{1.2}) is simply connected, and one has
corresponding covering maps
\be
\bR \times SQ(n) \longrightarrow U(1) \times SQ(n) \longrightarrow Q(n).
\label{1.3}\ee
The action-angle transforms \cite{RIMS95} of the alternative Sutherland phase spaces
\be
P:= T^*Q(n),
\quad
P_1:= T^*U(1) \times T^* SQ(n),
\quad
 P_2:= T^*\bR \times T^* SQ(n)
\label{1.4}\ee
are certain phase spaces
\be
\hat P_c,
\quad
\hat P_1:= T^*U(1) \times \bC^{n-1},
\quad
\hat P_2:= T^*\bR  \times \bC^{n-1}.
\label{1.5}\ee
The structure of $\hat P_c$  and the symplectic forms  are displayed in
equations (\ref{Pc}) and (\ref{4.33}) below,
but we note already here that $\hat P_c$ has the open dense submanifold
$\hat P$ defined by
\be
\hat P=\bT(n) \times \fC_x =\{ (e^{\ri \hat q}, \hat p)\},
\quad
\fC_x := \{ \hp \in \bR^n\,\vert\, \hp_i - \hp_{i+1} > \vert x \vert,
\quad i=1,\ldots, n-1\,\}
\label{1.6}\ee
equipped with the symplectic form $\hat \omega = \sum_{i=1}^n d\hat p_i \wedge d\hat q_i$
and the Hamiltonian
\be
\hat H_{\mathrm{RS}}(\hat q, \hat p)=
\sum_{a=1}^n (\cos \hat q_a) \prod_{k\neq a} \left[ 1 - \frac{x^2}{(\hp_k - \hp_a)^2}
\right]^{\frac{1}{2}},
\label{1.7}\ee
which is a particular real form of the complex rational Ruijsenaars-Schneider Hamiltonian \cite{RS}.
The flows of the commuting family of Liouville integrable Hamiltonians that contains $\hat H_{\mathrm{RS}}$
are complete only on the full phase space $\hat P_c$.

It will be shown  in  the present work that the relations between the above phase spaces,
described in \cite{RIMS95} without reference to Lie groups,
correspond to the covering homomorphisms
\be
G_2:=\bR \times SU(n) \longrightarrow G_1:=U(1) \times SU(n) \longrightarrow G:=U(n).
\label{1.8}\ee
For this, it will be used that the conjugation action of $G$ on $T^*G$ and the analogous
actions of $G_i$ on $T^*G_i$ ($i=1,2$) represent
actions of the same effective symmetry group
\be
\bar G:= G/\bZ_G \simeq G_1/ \bZ_{G_1} \simeq G_2/\bZ_{G_2},
\label{1.9}\ee
where $\bZ_G$ stands for the center of $G$ and so on.
By performing symplectic reductions at the usual
KKS value of the moment map of the $\bar G$-action, we shall obtain
 covering Poisson maps between the  respective reduced phase spaces,
\be
(T^* G_2)_{\red} \longrightarrow (T^*G_1)_{\red} \longrightarrow (T^*G)_\red,
\label{1.10}\ee
from the homomorphisms (\ref{1.8}).
By constructing two alternative models (symplectomorphic images) of each of the 3 reduced phase spaces in (\ref{1.10}) ,
 we shall arrive at the identifications
\be
P_2\simeq (T^* G_2)_\red \simeq \hat P_2,
\quad
P_1\simeq (T^* G_1)_\red \simeq \hat P_1,
\quad
P\simeq (T^* G)_\red \simeq \hat P_c.
\label{1.11}\ee
This will allow us to interpret the `duality symplectomorphisms'
\be
\cR_2: P_2 \longrightarrow \hat P_2,
\quad
\cR_1: P_1 \longrightarrow \hat P_1,
\quad
\cR: P \longrightarrow \hat P_c
\label{1.12}\ee
as the natural maps between the respective models of the same reduced phase space.
By virtue of (\ref{1.10}), then the following commutative diagram  arises
automatically from
our considerations:
\be
\xymatrix@C=6pc @R=1.5pc{
P_2
\ar[r]^-{{\mathcal R}_2}
\ar[d] &
\hat P_2 \ar[d]\\
P_1 \ar[r]^-{{\mathcal R}_1}
\ar[d]&
\hat P_1 \ar[d]\\
P\ar[r]^-{\mathcal R}&
{\hat P}_c }
\label{big}\ee
Here, the horizontal arrows represent duality symplectomorphims and the vertical arrows are
covering (locally symplectic) Poisson maps.
The maps corresponding to the arrows above have all  been constructed originally
by Ruijsenaars \cite{RIMS95}\footnote{To facilitate comparison, we note that our
 symbols $P$, $P_2$,  $\hat P_c$, $\hat P_2$, $\cR$, $\cR_2$ in (\ref{big})
correspond respectively to $\Omega$, $\tilde \Omega$, ${\hat \Omega}^\sharp$, ${\hat \Omega}^{\sharp c}$,
$\Phi^\sharp$, $\tilde \Phi^\sharp$ in diagram (1.74) of  \cite{RIMS95}, and our $\hat P$ in (\ref{1.6})
corresponds to $\hat \Omega$ in (1.67) of \cite{RIMS95}.},
relying on rather demanding direct arguments.
Our group theoretic interpretation throws a new light
on the web of dualities and coverings encapsulated by the above diagram.
This will also lead to technical simplifications in comparison to \cite{RIMS95}, concerning in particular
the symplectic property of the map $\cR$, which is obvious in our setting but was quite difficult to prove
originally by direct methods.

So far we have not  fully specified the commuting Hamiltonians of the dual pairs of systems, but it
will be explained in the main text that they  arise from two standard  families of
`free' Hamiltonians on $T^*G$, and from their pullbacks to $T^*G_i$, via the KKS reduction.
As for the logical organization of our arguments, we shall first describe the most complicated map
$\cR: P \to \hat P_c$ in detail, and then explain how the other maps in the diagram correspond
to the covering homomorphisms (\ref{1.8}).

In Section 2 we fix our notations and define the systems that we reduce subsequently.
We describe the two models $P$ and $\hat P_c$ of the reduced phase space $(T^*G)_\red$ in Section 3,
and derive also the dual pairs of Lax matrices that generate the commuting reduced Hamiltonians.
The structure of the maps in the diagram ({\ref{big})  is explained in Section 4.
Section 5 contains a brief summary together with a discussion of open problems.
An appendix is included, where we characterize the three alternative phase spaces (\ref{1.4}) of
`non-coinciding point-particles' moving on the line or on the circle.

\section{Setting the stage for symplectic reduction}
\setcounter{equation}{0}

Consider the Lie group $G=U(n)$ and identify its Lie algebra $\cG:= u(n)$ with the
dual space $\cG^*$ by means of the invariant scalar product
\be
\langle X, Y\rangle := -\operatorname{tr}(XY),
\qquad\forall X,Y\in \cG,
\label{2.1}\ee
and also use the identification
\be
T^*G \simeq G \times \cG^* \simeq G \times \cG= \{ (g,J)\,\vert\, g\in G,\, J\in \cG\}
\label{2.2}\ee
defined with the aid of right-translations on $G$. The cotangent bundle carries the symplectic form
\be
\Omega_{T^* G} = - d \operatorname{tr} (J dg g^{-1}).
\label{2.3}\ee
The group $\bar G$ (\ref{1.9}) can be realized as
$\bar G\simeq SU(n)/\bZ_n$ and this permits  identification of its Lie algebra
$\bar\cG$ with $su(n)$, which we regard also as  the model of $\bar\cG^*$ by taking advantage
of the same invariant scalar product as in (\ref{2.1}).
We then let $\cO \subset \bar\cG^*$ denote the minimal coadjoint orbit of $\bar G$
provided by
\be
\cO:= \{ \xi= \xi(x,v)\,\vert\, \xi(x,v):=
\ri x (\1_n - v v^\dagger ),\,\, v\in\bC^n,\,\,\vert v\vert^2= n \},
\label{2.4}\ee
where $x$ is a non-zero real parameter.
The Kirillov-Kostant-Souriau symplectic form $\Omega_\cO$ of $\cO$ can be written in
terms of the redundant variables  furnished by the components of the
vector $v$ of length $\sqrt{n}$  as
\be
\Omega_\cO =\ri x d v^\dagger \wedge dv.
\label{2.5}\ee
To clarify the meaning of formula (\ref{2.5}), we note that
the orbit $\cO$ coincides with the projective space $\bC P^{n-1}$ as
a manifold,  and we here view it as a reduction of the symplectic
vector space $\bC^n\simeq \bR^{2n}$ with respect to the natural action of $U(1)$
generated by the function  $v \mapsto \vert v\vert^2$.
By setting this $U(1)$ moment map equal
to the constant $n$ and then factoring by $U(1)$, the symplectic form of $\bC^n$, given by
$\ri x d v^\dagger \wedge dv$ but with unrestricted $v$,
 descends to the orbit $\cO$.
 The resulting symplectic form on $\cO$ is
a multiple of the standard Fubini-Study form of $\bC P^{n-1}$.

Our starting point for symplectic reduction will be the phase space
$(M, \Omega_M)$:
\be
M= T^* G \times \cO,
\qquad
\Omega_M = \Omega_{T^* G} + \Omega_{\cO}.
\label{2.6}\ee
Corresponding to the symplectic form $\Omega_M$, the non-vanishing Poisson brackets
between the matrix elements $g_{jk}$ and the components
$J_a:= \langle J, T_a\rangle$ and $\xi_a:= \langle \xi, T_a\rangle$ are
\be
\{ g_{jk}, J_a \}_M = (T_a g)_{jk},
\quad
\{ J_a, J_b \}_M = \langle J, [T_a,T_b]\rangle,
\quad
\{ \xi_a, \xi_b \}_M = \langle \xi, [T_a,T_b]\rangle,
\label{2.6+}\ee
where $\{ T_a\}$ is an arbitrary basis of $\cG$.

The phase space $M$ carries two families of  `free' Hamiltonians
$\{\cH_k\}$ and $\{\hat \cH_{\pm k}\}$ given by
\be
\cH_k(g,J,\xi):= \frac{1}{k}{\operatorname{tr}} (-\ri J)^k,
\quad
\forall k=1,\ldots, n,
\label{freeH}\ee
and
\be
\hat \cH_k(g,J,\xi) := \frac{1}{k}{\operatorname{tr}} ( g^k + g^{-k}),
\quad
\hat \cH_{-k}(g,J,\xi):=\frac{1}{\ri k } {\operatorname{tr}} (g^k - g^{-k}),
\quad
\forall k=1,\ldots, n.
\label{freehatH}\ee
Taking any initial value $(g(0), J(0), \xi(0))$,
the flow of the Hamiltonian $\cH_k$ can be written  as
\be
(g(t), J(t), \xi(t)) = (g(0)\exp(    \ri t (-\ri J(0))^{k-1} ), J(0), \xi(0)).
\label{2.9}\ee
For any positive integer $k$, the flow of the Hamiltonian $\hat \cH_{k}$  reads
\be
(g(t), J(t), \xi(t))= (g(0), J(0) + t (g(0)^k - g(0)^{-k}), \xi(0)),
\label{2.10}\ee
and the flow of $\hat \cH_{-k}$ is
\be
(g(t), J(t), \xi(t))= (g(0), J(0) -\ri t  (g(0)^k + g(0)^{-k}), \xi(0)).
\label{2.11}\ee
One has the Poisson brackets
\be
\{ \cH_k, \cH_l\}_M=0,
\qquad
\{ \hat \cH_a, \hat \cH_b\}_M=0
\label{2.12}\ee
for all possible integer subscripts that may occur.
In conclusion, the `spectral invariants' of $J$ and those of $g$ form
Abelian algebras of explicitly integrable Hamiltonians.

The `free' Hamiltonians are invariant under the effective action of
$\bar G$ on $M$ defined by assigning to each $[y]\in \bar G$ the symplectomorphism $A_{[y]}$ of $M$
that operates according to\footnote{The more correct notation
$A_{[y]}((g,J,\xi))$ is simplified to $A_{[y]}(g,J,\xi)$ throughout the paper.}
\be
A_{[y]}(g,J,\xi) := (y g y^{-1}, y J y^{-1}, y \xi y^{-1}),
\label{2.13}\ee
where
$y\in G$ is an arbitrary representative of $[y]\in  \bar G \simeq G/\bZ_G$.
This action is generated by the equivariant moment map
$\Phi: M \to \bar\cG^*\simeq su(n)$,
\be
\Phi(g,J,\xi) = J - g^{-1} J g + \xi.
\label{2.14}\ee
Indeed, $G$ acts by the
homomorphism $G\to \bar G$ and the vector field on $M$
corresponding to $T_a\in \cG$ is the Hamiltonian vector field of
$\Phi_a:= \langle \Phi, T_a\rangle$, as follows from (\ref{2.13}) and (\ref{2.6+}).
We are going to reduce at the value $\Phi=0$, which is a variant
of the KKS reduction \cite{KKS}.

It is known (and is easy to confirm along the lines indicated at the end of Subsection 3.1)
that zero is a regular value of the moment map
$\Phi$, and $\bar G$ acts freely on the constraint-manifold
\be
M_0:= \Phi^{-1}(0) \subset M.
\label{2.15}\ee
Therefore $M_0$ is an embedded submanifold of $M$ and the space of orbits
\be
(T^* G)_\red := M_0/\bar G
\label{2.16}\ee
is a smooth manifold.
This is the reduced phase space of the symplectic reduction of our interest.
The same reduced phase space can be obtained directly from $T^*G$ by `point reduction'
as well \cite{AbrMars},  but we shall find it convenient to proceed by utilizing the orbit $\cO$ as described
above.

The  reduction gives rise to the symplectic form $\Omega_\red$ and the
Abelian algebras of integrable reduced Hamiltonians $\{ H_k\}$ and $\{ \hat H_{\pm k}\}$ on
$(T^*G)_\red$ characterized by the equalities
\be
\pi_0^*(\Omega_\red) = \iota_0^* (\Omega_M),
\qquad
H_k \circ \pi_0 = \cH_k \circ \iota_0,
\qquad
\hat H_{\pm k} \circ \pi_0 = \hat \cH_{\pm k} \circ \iota_0,
\label{redHams}\ee
where $\pi_0: M_0 \to M_0/\bar G$ is the natural submersion and
$\iota_0: M_0 \to M$ is the tautological embedding.
The relations in (\ref{2.12}) imply  that
\be
\{ H_k, H_l\}_\red=0,
\qquad
\{ \hat H_a, \hat H_b\}_\red=0,
\ee
where the reduced Poisson bracket is associated with $\Omega_\red$.
The flows of the reduced Hamiltonians $H_k$ and $\hat H_{\pm k}$ result as the $\pi_0$-projections
of the flows of $\cH_k$ and $\hat\cH_{\pm k}$.

In symplectic reduction one often wishes to construct concrete models of the reduced phase space.
In principle, any symplectomorphic image of the reduced phase space can serve as a model  of it and
of course any two  models are automatically symplectomorphic to each other.
Speaking in terms of our specific example,  a \emph{global cross-section}
(also called \emph{global gauge slice}) is by definition
a submanifold  $N\subset M_0$ that intersects
every $\bar G$-orbit precisely once and is diffeomorphic to $M_0/\bar G$ by means
of the restriction of $\pi_0$.
A diffeomorphism between a submanifold $N\subset M_0$ intersecting
every $\bar G$-orbit precisely once and $M_0/\bar G$ is defined by the restriction of $\pi_0$
if and only if  the pull-back of $\Omega_M$ to  $N$ is symplectic.
Then  $N$ equipped with the pull-back of $\Omega_M$ is a model
of the reduced phase space $((T^*G)_\red, \Omega_\red)$.
In the subsequent Sections 3.1 and 3.2 we shall construct two models of our reduced phase space,
but only the second one will be obtained directly as a global gauge slice.

\section{Systems in duality from the reduction of $T^*U(n)$}
\setcounter{equation}{0}

In this section we present two models of the
reduced phase space $((T^*G)_\red, \Omega_\red)$.
The first model will be identified with the Sutherland phase space
$(P,\omega)$ (\ref{K.10}) since in terms of this model
the reduced Hamiltonians  $\{ H_k\}$ (\ref{redHams}) become the spectral invariants
of the Sutherland Lax matrix (\ref{K.11}).
The second model $(\hat P_c, \hat \omega_c)$ (\ref{Pc})
equipped with the commuting Hamiltonians $\{ \hat H_{\pm k}\}$ (\ref{redHams})
defines the Ruijsenaars dual of the Sutherland system.
It yields a completion of the rational Ruijsenaars-Schneider system $(\hat P, \hat \omega, \hat H_{\mathrm{RS}})$
described in the Introduction.
The identification of $((T^*G)_\red,\Omega_\red)$ with the Sutherland phase space
$(P,\omega)$ is well-known \cite{KKS}.
The identification of $((T^*G)_\red, \Omega_\red)$
with $(\hat P_c, \hat \omega_c)$ is constructed by merging the methods
applied in the previous papers \cite{FK-JPA,FK3}. In essence, the two models represent two coordinate systems
on $(T^* G)_\red$, and the change of coordinates gives the duality map (action-angle transform)
of Ruijsenaars \cite{RIMS95} as we explain at the end of the section.

\subsection{The KKS derivation of the Sutherland system}

Let $\bT(n)^0$ denote the regular part of the maximal torus
\be
\bT(n) = \underbrace{U(1)\times  U(1) \times \cdots \times U(1)}_{n-\mathrm{times}} < U(n).
\label{K.2}\ee
The open submanifold $\bT(n)^0 \subset \bT(n)$, realized as
\be
\bT(n)^0 =\{ \tau =\diag(\tau_1,\ldots, \tau_n)\in \bT(n)\,\vert\, \tau_a\neq \tau_b
\quad \hbox{for all}\quad 1\leq a\neq b \leq n\},
\label{K.1}\ee
is the configuration space of $n$ distinguished `non-coinciding point-particles' moving on the circle
$U(1)$.
The permutation group $S(n)$ acts freely on $\bT(n)^0$, by permuting the entries of
$\tau$, and therefore the space of orbits
\be
Q(n):= \bT(n)^0/S(n)
\label{K3}\ee
is a smooth manifold, such that the natural projection
$\bT(n)^0 \to Q(n)$ is a smooth submersion.
We note (see Appendix A) that $\bT(n)^0$ has $(n-1)!$ connected components and $Q(n)$ is
connected.
By definition, \emph{$Q(n)$ is the configuration space of $n$ indistinguishable particles
moving on the circle}.

The cotangent bundle $(T^* \bT(n)^0, \Omega_{T^* \bT(n)^0})$ can be identified as
\be
T^*\bT(n)^0= \bT(n)^0 \times \bR^n=
\{ (\tau,  p),\,\vert\, \tau \in \bT(n)^0,\,p\in \bR^n\},
\quad
\Omega_{T^*\bT(n)^0} = \sum_{k=1}^n d p_k \wedge \frac{d\tau_k}{\ri \tau_k}.
\label{K.4}\ee
If we use the parametrization $\tau_k = e^{\ri q_k}$ with local coordinates $q_k$, then
the symplectic form takes the usual Darboux form
\be
\Omega_{T^* \bT(n)^0}= \sum_{k=1}^n d p_k \wedge dq_k.
\label{K.5}\ee
The permutation group $S(n)$ acts freely  also  on $T^*\bT(n)^0$ by the cotangent lift of its
action on $\bT(n)^0$, and it follows from well-known general results \cite{AbrMars} that
the corresponding space of $S(n)$-orbits is the cotangent bundle of $Q(n)$:
\be
( T^*Q(n), \Omega_{T^* Q(n)}) = (T^* \bT(n)^0, \Omega_{T^* \bT(n)^0})/S(n).
\label{K.6}\ee
The projection
$
\bT(n)^0 \to Q(n)
$
is locally a diffeomorphism (a covering) and it induces the map (another covering)
$
T^* \bT(n)^0 \to T^* Q(n)
$,
whereby the pull-back of $\Omega_{T^* Q(n)}$ equals $\Omega_{T^* \bT(n)^0}$.
In the coordinates $(\tau, p)$ on $T^* \bT(n)^0$ (\ref{K.4}) the cotangent lift of
$\sigma \in S(n)$ acts simply as
\be
\sigma: (\tau, p) \mapsto (\sigma(\tau), \sigma(p)),
\label{K.9}\ee
where $\sigma(\tau)$ and $\sigma(p)$ are obtained by applying the permutation $\sigma$
to the entries of $\tau$ and $p$.
In short, \emph{one may regard $T^*Q(n)$ as $T^* \bT(n)^0$ where the components of $\tau$ and $p$
matter only up to simultaneous permutations}.
As we explain in Appendix A,
$Q(n)$ has a non-trivial topological structure.
Just because of that non-trivial structure, it is often advantageous to
replace $Q(n)$ by $\bT(n)^0$ with the proviso that `everything matters up to
permutations'.
In particular, we may identify the smooth functions on $T^* Q(n)$ with the smooth
$S(n)$-invariant functions on $T^* \bT(n)^0$.

\medskip
\noindent
{\bf Definition 3.1.}
The Sutherland system as a system of  indistinguishable particles possesses the phase space
\be
(P, \omega):= (T^*Q(n), \Omega_{T^* Q(n)})
\label{K.10}\ee
equipped with the commuting Hamiltonians given by the spectral invariants of the Lax matrix
\be
L_{\mathrm{Suth}}(q,p)  :=
\sum_{k=1}^n p_k E_{kk} - \frac{\ri x}{2}\sum_{a\neq b} \frac{E_{ab}}{\sin\frac{q_a - q_b}{2}}.
\label{K.11}\ee
Here $L_{\mathrm{Suth}}(q,p)$  is viewed as a function on $T^* \bT(n)^0$ (\ref{K.4}),
with $\tau = e^{\ri q}$,
and its symmetric functions yield functions on $T^* Q(n)$ due to their $S(n)$-invariance.
Throughout the paper,  $E_{ab}$ denotes the $n\times n$  matrix having
a single non-zero element, equal to $1$, at the $ab$ position.

\medskip
The following results about the reduction of $(M, \Omega_M)$ (\ref{2.6})
are due to Kazhdan, Kostant and Sternberg \cite{KKS}.
We present them together with a proof for the sake of readability.

\medskip
\noindent
{\bf Theorem 3.2.}
\emph{
The image of the smooth, injective map $F: T^*\bT(n)^0 \to M$ defined by
\be
F: (\tau, p) \mapsto (\tau, J(\tau,p), \xi(x,\hat v)),
\quad
 J(\tau,p):=\ri \sum_{k=1}^n p_k E_{kk} + \ri  \sum_{a\neq b} \frac{x E_{ab}}{ 1 - \tau_b/\tau_a},
\quad
\hat v:= [1,1,\ldots, 1]^T
\label{K.12}\ee
is an embedded submanifold $M^F \subset M_0$ (\ref{2.15}) that intersects every gauge orbit.
If a gauge transformation by $[y]\in \bar G$ maps a point of $M^F$ into $M^F$, then
its representative $y\in G$ can be taken to be a permutation matrix.
Every permutation matrix $\sigma\in G$ maps $M^F$ to $M^F$, and $F$ is an $S(n)$-equivariant map
with respect to the actions of $S(n)$ on $T^* \bT(n)^0$ and on $M^F$. Finally, there holds the relation
\be
F^* (\Omega_M)= \Omega_{T^* \bT(n)^0} .
\label{K.13}\ee
}
\medskip
\noindent
{\bf Corollary 3.3.} \emph{The cotangent bundle $(T^* Q(n), \Omega_{T^* Q(n)})$ is a model of the
reduced phase space $((T^*G)_\red,\Omega_\red)$ (\ref{2.16}).
If $(\tau,p)$ is  a representative of
$[(\tau,p)]\in T^* Q(n) = T^* \bT(n)^0/ S(n)$,
and $\pi_0: M_0 \to (T^*G)_\red$ is the projection, then
a symplectomorphism $\cF: T^* Q(n) \to (T^*G)_\red$ is given by
\be
\cF: [(\tau,p)] \mapsto (\pi_0 \circ F)(\tau,p).
\label{K.15}\ee
By using this symplectomorphism and the identification of the  functions on
$T^* Q(n)$ with $S(n)$-invariant functions on $T^* \bT(n)^0$,
the reduced Hamiltonians $H_k$  (\ref{redHams})  take the form
\be
H_k(\tau,p) =\frac{1}{k} {\operatorname{tr}} (-\ri J(\tau, p))^k,
\qquad k=1,\ldots,n,
\ee
and the reduced Hamiltonians $\{ \hat H_{\pm k}\}$ (\ref{redHams}) are furnished by
\be
\hat H_{k}(\tau,p) =\frac{1}{k} \sum_{j=1}^n ((\tau_j)^{k} + (\tau_j)^{-k}),
\qquad
\hat H_{- k}(\tau,p) =\frac{1}{k\ri } \sum_{j=1}^n ((\tau_j)^{k} - (\tau_j)^{-k}).
\label{K.16}\ee
 By setting $\tau_j = e^{\ri q_j}$,
the functions $H_k$ become  spectral invariants of the Lax matrix $L_{\mathrm{Suth}}(q,p)$.
}

\medskip
\noindent
{\bf Proof.}
We have to check the validity of the middle two  in the following chain of identifications:
\be
(T^*G)_\red = M_0/\bar G \simeq M^F/S(n)  \simeq T^* \bT(n)^0/S(n) = T^* Q(n).
\label{K.18}\ee
We start by recalling that the moment map constraint is
\be
J - g^{-1} J g +\xi(x,v)=0\quad\hbox{with}\quad  \xi(x,v)= \ri x (  \1_n - v v^\dagger),
\quad
\vert v \vert^2 =n.
\label{K.19}\ee
By using the gauge freedom, we can transform any solution of this constraint into a solution
for which $g$ belongs to the maximal torus. That is, it is
sufficient to solve the constraint
\be
J - \tau^{-1} J \tau = \ri x ( v v^\dagger - \1_n),
\qquad
\vert v \vert^2 =n,
\qquad
\tau\in \bT(n).
\label{K.20}\ee
We can see from the diagonal part of this equation that the diagonal components of
$J$ are arbitrary, and also see that
$\vert v_k \vert =1$
for each $k=1,\ldots, n$.
Therefore we can bring the solution, by a gauge transformation
by an element of
$\bT(n)$, into a solution for which $v= \hat v$ defined in (\ref{K.12}).
Then the off-diagonal components of the constraint become
\be
J_{ab}( 1- \tau_b/\tau_a) = \ri x,
\qquad
\forall a\neq b.
\label{K.23}\ee
This can be solved if and only of $\tau \in \bT(n)^0$ (\ref{K.1}) ,
and the solution is given precisely by the formula $J(\tau,p)$ in (\ref{K.12}).
To summarize, so far we have shown that
\be
(g,J, \xi(x,v)) = (\tau, J(\tau,p), \xi(x,\hat v))
\label{K.24}\ee
is a solution of the moment map constraint for all $\tau \in \bT(n)^0$ and $p\in \bR^n$,
and every solution is a gauge transform of a solution of this form.
Notice that these solutions form precisely the image $M^F$ of the map $F$ (\ref{K.12}).
Note also that
\be
L_{\mathrm{Suth}}(q,p) = -\ri e^{-\ri q/2} J(e^{\ri q}, p) e^{\ri q/2}.
\label{K.25}\ee

Consider now a `residual gauge transformation' $[y]\in \bar G$ that maps a solution
of the above form into another (or the same) solution of the above form, i.e.,
\be
(y \tau y^{-1}, y J(\tau, p) y^{-1}, \xi(x, y \hat v))
= (\tau', J(\tau', p'), \xi(x, \hat v)).
\label{K.26}\ee
We conclude from the equality $y \tau y^{-1} = \tau'$ that $y$ must have the form
$
y = \sigma T
$,
where $\sigma\in G$ is a permutation matrix and $T\in \bT(n)$. Then we infer from the
third component of (\ref{K.26}) that $T$ must be a multiple of the unit matrix.
Returning to the first and second components, we see that
\be
\tau' = \sigma \tau \sigma^{-1} \equiv \sigma(\tau)
\quad\hbox{and}\quad
\sum_k p'_k E_{kk} = \sigma (\sum_k p_k E_{kk}) \sigma^{-1} \equiv \sum_{k=1}^n \sigma(p)_k E_{kk}.
\label{K.28}\ee
This shows that $F$ is an $S(n)$-equivariant bijection between $T^* \bT(n)^0$ and $M^F$,
and it is also easily checked that  $F^* (\Omega_M) = \Omega_{T^* \bT(n)^0}$ holds.
Now all statements of the theorem and the corollary follow immediately from
the properties of $F$ and its image $M^F$ that we have established, and from (\ref{K.25}).
Incidentally, it is also clear from the above  that the isotropy subgroup of
any $(\tau, J(\tau, p), \xi(x, \hat v))\in M^F$ is the trivial subgroup of $\bar G$, which
 entails that
$\bar G$ acts freely on $M_0$.
\hspace*{\stretch{1}} \qedsymb

\medskip

For completeness,
the reader may wish to verify that
zero is a regular value
of the moment map $\Phi$, i.e.,
 the derivative map $D\Phi(g,J,\xi): T_{(g,J,\xi)} M \to su(n)$
is surjective at every point of $M_0=\Phi^{-1}(0)$.
The required inspection is readily performed at the points of $M^F\subset M_0$, which
is sufficient since $\Phi$ is equivariant and $M^F$ intersects every gauge orbit in
$M_0$.

\subsection{Derivation of the Ruijsenaars dual of the Sutherland system}

We have seen that  the Sutherland phase space $(P,\omega)$ (\ref{K.10}) is a model
of $(T^*G)_\red$ defined by the KKS reduction of $(M,\Omega_M)$.
Now our aim is to construct a `dual' model of the reduced phase space $(T^*G)_\red$.
For this purpose we devise an alternative way to solve
the moment map constraint (\ref{K.19}).
In the preceding subsection we proceeded by diagonalizing $g\in G$, and here
we start from the observation that
every solution of (\ref{K.19}) can be obtained as a gauge transform
of a solution for which $J\in \cG$ is diagonal of the form
\be
 J= -\ri\, \diag(\hp_1, \ldots, \hp_n) := - \ri \hp
\quad\hbox{with}\quad
\hp_1 \geq \hp_2 \geq \cdots \geq \hp_n.
\label{a6}\ee
The final result, given by Theorem 3.12, will be
reached through a series of auxiliary lemmas.

\noindent
{\bf Lemma 3.4.}
\emph{
If $(g,J, \xi(x,v))$ is a solution of the moment moment map constraint (\ref{K.19})  with $J$ in (\ref{a6})
then
\be
\hp_k - \hp_{k+1} \geq \vert x \vert,
\qquad \forall k=1,\ldots, (n-1),
\label{a7}\ee
and
\be
\vert v_b\vert^2 = \prod_{k\neq b}
\left[ \frac{\hp_b - \hp_k - x}{\hp_b - \hp_k}\right],
\qquad
\forall b=1,\ldots, n.
\label{a8}\ee
}

\noindent
{\bf Proof.}
We  rewrite the moment map constraint (\ref{K.19}) in the equivalent form
\be
 g^{-1} \hp  g =  x (v v^\dagger - \1_n) + \hp.
\label{a9}\ee
We can compute the characteristic polynomials of the matrices on the two sides
of this equation. In this way we deduce the equality of the polynomials
\be
\prod_{j=1}^n (\hp_j - \lambda) =
\prod_{j=1}^n (\hp_j - (\lambda +x)) +
x \sum_{k=1}^n \bigl(\vert v_k \vert^2 \prod_{j\neq k} (\hp_j - (\lambda +x))\bigr).
\label{a10}\ee

Suppose now that $\hp$ is regular, i.e.,
\be
\hp_1 > \hp_2 > \cdots > \hp_n.
\label{a11}\ee
For regular $\hp$ (\ref{a11}),  the evaluation of (\ref{a10})  at $\lambda = \hp_b - x$ gives immediately
the relation (\ref{a8}).
By using (\ref{a8}) together with (\ref{a11}) and $\vert v_b \vert^2 \geq 0$, it is not
difficult to obtain the `spectral gap condition' (\ref{a7}).
Indeed, one may follow the argumentation presented in \cite{FK3} in connection with a completely analogous problem.

Let us continue by showing that solutions of (\ref{a9}) satisfying (\ref{a11}) exist.
In fact, one may take the explicit example defined for any $\hp$ with
\be
\hp_k - \hp_{k+1}= \vert x \vert,
\qquad
k=1,\ldots, n-1,
\label{a12}\ee
as follows:
\be
v_i = \delta_{n,i} \sqrt{n},
\quad
g_{1,n}= g_{j, j-1}=1\quad (j=2,\ldots, n), \quad g_{a,b}=0 \quad \hbox{otherwise,}\quad
\hbox{if}\quad x>0,
\label{a13}\ee
and
\be
v_i = \delta_{1,i} \sqrt{n},
\quad
g_{n,1}= g_{j, j+1}=1\quad (j=1,\ldots, n-1),
\quad g_{a,b}=0 \quad \hbox{otherwise,}\quad\hbox{if}\quad x<0.
\label{a14}\ee

We observe from the foregoing arguments that for all solutions of (\ref{K.19}) for which the eigenvalues
of $J$ are all distinct the distance of any two eigenvalues of $J$ is at least $\vert x\vert$,
and such regular solutions do exist.
This allows us to conclude that there cannot be any solution for which two
eigenvalues of $J$ coincide. Indeed, we know that the constraint-manifold $M_0$ is connected.
(Recall from Section 3.1 that
$M_0$ is a principal fiber bundle with connected fiber $\bar G$ and connected base $T^* Q(n)$.)
Therefore any hypothetical non-regular solution would be continuously connected to
a regular solution. However, this contradicts the lower bound $\vert x\vert$ in the distance
of the eigenvalues of any regular solution.
\hspace*{\stretch{1}} \qedsymb

\medskip
\noindent
{\bf Definition 3.5.}
Let $\bar \fC_x$ denote the closure of  the `Weyl chamber with thick walls'
$\fC_x$ introduced in (\ref{1.6}).
Define the function $V(x): \bar\fC_x\to \bR^n$  by the formula
\be
V(x,\hp)_b:=
\prod_{k\neq b}
\left[ \frac{\hp_b - \hp_k - x}{\hp_b - \hp_k}\right]^{\frac{1}{2}},
\qquad
\forall b=1,\ldots, n,
\label{a17}\ee
and introduce the real $n\times n$ matrix valued smooth function
$\eta(x,\hp)$ on $\bar\fC_x$ by the formula
\be
\eta(x,\hp)_{ab}= \frac{x}{ \hp_b - \hp_a} \prod_{j\neq a,b}\left[
\frac{ (\hp_a - \hp_j - x) (\hp_j - \hp_b - x)}
{ (\hp_a - \hp_j) (\hp_j - \hp_b)}\right]^{\frac{1}{2}},\quad
\quad a\neq b,
\label{a20}\ee
\be
\eta(x,\hp)_{aa} = \prod_{j\neq a}\left[
\frac{ (\hp_a - \hp_j - x) (\hp_j - \hp_a - x)}
{ (\hp_a - \hp_j) (\hp_j - \hp_a) }\right]^{\frac{1}{2}}.
\label{a21}\ee
Note that the expression under the square root is non-negative in each factor in the above three
equations and the non-negative square root is taken.

\medskip
\noindent
{\bf Lemma 3.6.}
\emph{There exists a solution of the moment map constraint (\ref{K.19}) of the form
\be
(g,J,\xi)= (g, - \ri \hp, \xi(x, V(x, \hp)))
\ee
for every $\hp \in \bar\fC_x$. Here, we use the notation
$\hat p= \diag(\hp_1,\ldots, \hp_n)$. }

\noindent
{\bf Proof.}
Let us arbitrarily fix $\hp \in \bar\fC_x$.
Recall that the moment map constraint requires the existence of $g\in U(n)$ for which
\be
 g^{-1} \hp   g =  x (V(x,\hp) V(x,\hp)^\dagger - \1_n) + \hp.
\ee
Since the matrix on the right-hand-side is Hermitian, the existence of such a $g$ is guaranteed
if we show that the characteristic polynomial $\cQ_n(\lambda, \hp)$ of the matrix
on right hand side is equal to $\cP_n(\lambda):= \prod_{j=1}^n (\hp_j - \lambda)$.
The very definition of $V(x,\hp)$ (see the argument after (\ref{a11})) guarantees that
\be
(\cP_n-\cQ_n)(\lambda= \hp_b -x,\hp) =0,
\qquad
\forall b=1,\ldots, n.
\ee
On the other hand, it is obvious from their definition that $\cP_n-\cQ_n$ is a polynomial in $\lambda$ of degree
strictly lower than $n$. Therefore we have the equality $\cP_n(\lambda, \hp)= \cQ_n(\lambda, \hp)$.
\hspace*{\stretch{1}} \qedsymb

\noindent
{\bf Lemma 3.7.}
\emph{
The function $V(x,\hp)$ (\ref{a17}) satisfies the identities
\be
\sum_{a=1}^n \frac{x}{\hp_b - \hp_a + x} V(x, \hp)_a^2  =1,
\quad
\forall b=1,\ldots, n, \quad \forall \hp \in \fC_x,
\label{a18}\ee
\be
\sum_{a=1}^n V(x, \hp)_a^2 = n,
\qquad
\forall \hp\in \bar\fC_x.
\label{a19}\ee
The function $\eta(x,\hp)$ enjoys the properties
\be
\eta(x, \hp)_{ab}=\frac{ x V(x,\hp)_a V(-x,\hp)_b}{ \hp_b - \hp_a + x},
\qquad
\forall a,b\quad \hbox{if} \quad \hp\in \fC_x,
\label{3.40}\ee
\be
\eta(x,\hp)^{-1} = \eta(x, \hp)^T = \eta(-x, \hp),
\quad
\det (\eta(x,\hp))=1,
\qquad \forall \hp\in \bar\fC_x.
\label{a22}\ee
}

\noindent{\bf Proof.}
The identity (\ref{a19}) follows by taking the trace of (\ref{a9}), and (\ref{a18})
follows by evaluation of the equality
of the characteristic polynomials (\ref{a10}) at $\lambda=\lambda_b$.
We note in passing that the identity (\ref{a18}) also extends smoothly to the closure
$\bar \fC_x$ since the singularities coming from the denominators cancel against the zeros
of the components of $V$.
Regarding $\eta(x,\hp)$,
the only non-trivial statements are the first equality in (\ref{a22}) and the claim about the determinant being $1$.
These statements follow from (\ref{3.40}) by using the Cauchy determinant formula (see e.g.~\cite{Pras})
and the continuity of $\eta$ on $\bar\fC_x$.
\hspace*{\stretch{1}} \qedsymb

Note that the functions $V(x,\hp)$ and $\eta(x,\hp)$ as well as their
properties given by Lemma 3.7 can be found in \cite{RIMS95}, too.
Observe from (\ref{a22})  that $\eta(x,\hat p)\in SO(n,\bR) < SU(n)$.

\medskip
\noindent
{\bf Lemma 3.8.}
\emph{By using the above notations,
the following formula defines a solution of the moment map constraint (\ref{K.19}) for each $\hp \in \bar\fC_x$:
\be
(g, J, \xi) = (\eta(x,\hp)^{-1}, -\ri \hp, \xi(x,V(x,\hp))).
\label{a23}\ee
}

\medskip
\noindent
{\bf Proof.}
By multiplying (\ref{a9}) by $g^{-1}$ from the right and substituting (\ref{a23}), we see that the
statement is equivalent to the identity
\be
\eta(x,\hp)_{ab} ( \hp_b - \hp_a  + x)
= x V_a(x,\hp) \sum_{c=1}^n V_c(x,\hp) \eta(x, \hp)_{cb},
\qquad \forall a,b.
\ee
By using the above formulae, it is readily verified
that both sides are equal to
\be
x V(x, \hp)_a V(-x,\hp)_b.
\ee
The easiest way is to first check this identity on $\fC_x$, and then notice
that continuity guarantees that it remains valid also on the closure $\bar\fC_x$.
\hspace*{\stretch{1}} \qedsymb

\medskip
\noindent
{\bf Lemma 3.9.} \emph{Applying the notations of (\ref{2.13}) and Definition 3.5, define the continuous map
\be
K_x: \bar G  \times \bT(n) \times \bar \fC_x \to M\simeq U(n) \times u(n) \times \cO
\label{3.45}\ee
by the formula
\be
K_x([y], \cD, \hp) = A_{[y]}\!\left( \left(\eta(x, \hp) \cD \right)^{-1} , - \ri  \hp ,  \xi(x,V(x,\hp)) \right).
\label{3.46}\ee
Then the image of $K_x$ coincides with the constraint-manifold $M_0$ (\ref{2.15}).
The restriction of  $K_x$ to $\bar G \times \bT(n) \times  \fC_x$
is smooth, injective and its image is a dense, open submanifold of $M_0$.
}

\medskip
\noindent
{\bf Proof.}
It follows from Lemmas 3.4 and 3.6 that every element of $M_0$ can be obtained as a gauge transform of
an element of $M_0$ of the form
\be
(g, -\ri \hp, \xi(x, V(x, \hp)))
\quad\hbox{with some}\quad \hp \in \bar\fC_x.
\ee
In this case the moment map constraint amounts to the following equation for $g\in U(n)$:
\be
\ri g^{-1} \hp g = - \xi(x, V(x,\hp)) + \ri \hp.
\label{a37}\ee
By noting that $\hp$ is regular and that we have Lemma 3.8,
we see that
(\ref{a37}) is solved if and only if
\be
g=  \left(\eta(x, \hp) \cD\right)^{-1}
\quad\hbox{for some}\quad \cD \in \bT(n).
\ee
These arguments show that the image of $K_x$ is indeed $M_0$.
It is also obvious that $K_x$ maps  $\bar G \times \bT(n) \times \fC_x$
onto a dense, open submanifold of $M_0$.
The restricted map is smooth on account of its formula, and
to show that it is injective suppose that
\be
A_{[y]} \bigl(\left(\eta(x, \hp)\cD\right)^{-1}, - \ri \hp , \xi(x,V(x,\hp)) \bigr)
= A_{[w]}\bigl(\left(\eta(x, \hp')\cD'\right)^{-1}, - \ri  \hp' , \xi(x,V(x,\hp'))\bigr)
\label{eq3}\ee
for some
\be
([y], \cD, \hp) \in \bar G \times \bT(n) \times \fC_x \ni ([w], \cD', \hp').
\ee
Comparison of the second component of the triples in (\ref{eq3}) implies that $\hp = \hp'$ and
\be
g_0:=w^{-1} y\in \bT(n).
\ee
Moreover, it then follows from the third component of (\ref{eq3}) that
\be
g_0 \xi(x, V(x, \hp)) g_0^{-1} = \xi(x, g_0 V(x, \hp)) = \xi(x, V(x, \hp)).
\ee
This means that
\be
g_0 V(x, \hp) = \gamma V(x, \hat p)
\quad
\hbox{for some}\quad \gamma \in U(1).
\ee
Taking into account that all components of $V(x, \hp)$ are non-zero, since $\hp \in \fC_x$, we get that
$
g_0= \gamma \1_n.
$
The statement now follows by looking at the first component of
the equality (\ref{eq3}).
\hspace*{\stretch{1}} \qedsymb

\medskip

Define
\be
M_0^0 \subset M_0
\label{3.55}\ee
to be the set of the elements
$(g,J,\xi)\in M_0$ for which the eigenvalues of $J$ satisfy the strict spectral gap condition, i.e.,
for which $\ri J$ is conjugate to an element of $\fC_x$.
It is clear that $M_0^0 \subset M_0$ is a dense, open, $\bar G$-stable submanifold, which
is  the image of $\bar G \times \bT(n) \times \fC_x$ by the map $K_x$.
Correspondingly,
\be
M_0^0/ {\bar G} \subset M_0/{\bar G}
\label{a34}\ee
is a dense, open submanifold of the reduced phase space.

\medskip
\noindent
{\bf Lemma 3.10.}
\emph{
Consider the smooth, injective map $m_x: \bT(n) \times \fC_x \to M$ given by
\be
m_x(\cD, \hp) := K_x( [e], \cD, \hat p)=
( \left(\eta(x, \hp)\cD\right)^{-1} , - \ri \hp ,  \xi(x,V(x,\hp))).
\ee
The image of $m_x$ lies in $M_0^0$ and it intersects every gauge orbit in $M_0^0$ precisely once.
Moreover,  $m_x$ pulls-back the symplectic form $\Omega_M$ of $M$ (\ref{2.6}) according to
\be
m_x^*(\Omega_M) = -\ri \operatorname{tr}\left(d \hp \wedge (d \cD)\cD^{-1}\right).
\label{kx*}\ee
}

\noindent
{\bf Proof.}
The only task is to verify (\ref{kx*}).
Indeed, the smoothness of $m_x$
is obvious from its formula and it enjoys the properties mentioned above as an
immediate consequence of Lemma 3.9.
To verify (\ref{kx*}) we put  $g:= \left( \eta(x,\hp)\cD\right)^{-1}$ and $J = -\ri \hp$,
which  satisfy the equality
\be
{\operatorname{tr}}\left( J (d g) g^{-1}\right)   =   {\operatorname{tr}} \left(\ri \hp (d \cD)\cD^{-1}\right).
\ee
The basic fact that implies this is that $\eta$ is a real orthogonal matrix, and
thus it gives no cross term with $\hp$.
It is also important  to notice that the `orbital part' $\Omega_{\cO}$ of the
symplectic form $\Omega_M$  gives zero contribution to $m_x^* (\Omega_M)$.
This follows from the formula (\ref{2.5}) of the symplectic form $\Omega_\cO$
in terms of the redundant variable $v$, as it
becomes zero upon restriction to any submanifold consisting of vectors
$v\in \bC^n$ with purely real components.
Since the components of $V(x,\hp)$ are all real, the statement (\ref{kx*}) follows.
\hspace*{\stretch{1}} \qedsymb

 For each $\tau=\diag(\tau_1,\dots,\tau_n)\in\bT(n)$  set
\be
\tau_{(x)}:=\diag(\tau_2,\dots,\tau_{n},1) \quad\hbox{if}\quad  x>0,  \qquad
\tau_{(x)}:=\diag(1,\tau_1,\dots,\tau_{n-1})\quad\hbox{if}\quad x<0.
\label{defoftx}\ee
Introduce also  the bijection
 $\aleph(x,\cdot ): \bT(n) \to \bT(n)$  by
\be
\aleph(x,\tau)_j:=\prod_{k=j}^n\tau^{-1}_k, \quad x>0
\quad\hbox{and}\quad
  \aleph(x,\tau)_j:=\prod_{k=1}^j\tau^{-1}_k, \quad x<0.
\label{deff}\ee
Note  the identity
\be
 \aleph(x,\tau)(\aleph(x,\tau))_{(x)}^{-1} = \tau^{-1},
 \qquad
 \forall \tau\in \bT(n).
 \label{alephprop}\ee

We described in the Introduction the Hamiltonian system
$(\hat P, \hat \omega, {\hat H}_{\mathrm{RS}})$, which is
a real form of the complex rational Ruijsenaars-Schneider system.
The commuting Hamiltonians of this system are the spectral invariants
of the Lax matrix
\be
\hat L(\hat q, \hat p) :=
\eta(x,\hat p) e^{\ri \hat q},
\label{LRS}\ee
 viewed as a function on $\hat P$ (\ref{1.6}). In particular,  the Hamiltonian (\ref{1.7}) obeys
 \be
 \hat H_{\mathrm{RS}}(\hat q, \hat p) =
 \frac{1}{2} {\operatorname{tr}}\bigl(\hat L(\hat q, \hat p) + \hat L(\hat q, \hat p)^{-1}\bigr).
 \label{HRS}\ee

The usefulness
of the next reformulation of Lemma 3.10
will be justified by the final result.

\medskip
\noindent
{\bf Proposition 3.11.}
\emph{
With $(\hat P, \hat \omega)$  in (\ref{1.6}) and $K_x$ in (\ref{3.46}),
the map $k_x: \hat P \to M$ defined by
\be
k_x(e^{\ri \hat q}, \hp):= K_x( [\aleph(x, e^{\ri \hat q})_{(x)}], e^{\ri \hat q}, \hat p)
\label{kx}\ee
enjoys the same properties as the map $m_x$ of Lemma 3.10.
In particular, $k_x^*(\Omega_M) = \hat \omega$.
Thus the symplectic manifold $(\hat P, \hat \omega)$ provides a model of the
dense, open submanifold $M_0^0/\bar G$ of the reduced phase space.
On this dense open submanifold the Abelian family $\{ \hat \cH_{\pm k}\}$ (\ref{freehatH}) reduces to
the spectral invariants of the Lax matrix $\hat L(\hat q, \hat p)$ (\ref{LRS}), and the other
Abelian family $\{ \H_k\}$ (\ref{freeH}) reduces to the symmetric
polynomials of the `dual position variable' $\hat p$.}

\medskip\noindent
{\bf Proof.}
The claims follow from Lemma 3.10 since the maps $m_x$ and $k_x$ are
gauge equivalent.
\hspace*{\stretch{1}} \qedsymb

The results presented above characterize the dense submanifold $M_0^0/\bar G \subset M_0/\bar G$
in `dual variables'.
For a full description, we shall construct
a global cross-section of the gauge orbits in the complete constraint-manifold $M_0$.
This will be achieved by constructing an extension of the map $k_x$ (in the sense of Eq.~(\ref{prop}) below).
To save place, in the rest of this section we  assume that
\be
x>0.
\ee

We introduce the symplectic manifold $(\hat P_c, \hat \omega_c)$ by setting
\be
\hat P_c:=   \bC^{n-1} \times  \bC^\times,
\qquad \hat\omega_c:=
\frac{\ri dZ\wedge d\bar Z}{2\bar ZZ} +\sum_{j=1}^{n-1}\ri dz_j\wedge d\bar z_j,
\quad Z\in \bC^\times, \quad z\in \bC^{n-1},
\label{Pc}\ee
where  $\bC^\times$  denotes
 the complex plane without the origin.
 Following \cite{RIMS95,FK3},
 we then define the  smooth, injective map
  $\cZ_x: \hat P\equiv \bT(n) \times \fC_x \to \hat P_c$  by the  formulae
 \be
 z_j(e^{\ri \hat q},\hat p):=(\hat p_j-\hat p_{j+1}-x)^{\frac{1}{2}}\prod_{k=j+1}^ne^{-\ri\hat q_k},
 \,\, j=1,..., n-1, \,\, Z(e^{\ri \hat q},\hat p):=e^{-\hat p_1} \prod_{k=1}^ne^{-\ri\hat q_k},
  \label{x+}\ee
and let $\tilde \cZ_x: \bT(n) \times \bar\fC_x \to \hat P_c$
stand for the unique continuous extension of $\cZ_x$.
The component-functions of $\tilde \cZ_x$ are given by the same formulae as those of $\cZ_x$.
The map $\tilde \cZ_x$ is surjective, but is obviously not injective.
The image
\be
\hat P_c^0:= \cZ_x(\hat P)  \subset \hat P_c
\label{3.69}\ee
is the dense, open submanifold consisting of the points for which  $z_j\neq 0$ for all $j$.
We have
\be
\cZ_x^* (\hat \omega_c) = \sum_{i=1}^n d\hp_i \wedge d\hat q_i \equiv \hat \omega.
\label{59}\ee
This means that $\cZ_x$ yields a symplectomorphism between
$(\hat P, \hat \omega)$ and $(\hat P_c^0, \hat \omega_c)$.

Now we introduce the functions $\hat \pi_j(z,Z)$, the matrix function
$\vartheta(z,Z)$, and the vector-function $\cV(z,Z)$ on $\hat P_c$ by the following
defining formulae:
\bea
&&\hat \pi_j(z(e^{\ri \hat q},  \hp), Z(e^{\ri \hat q},  \hp))  = \hp_j, \\
&&\cV(z(e^{\ri \hat q},  \hp), Z(e^{\ri \hat q},  \hp))  = \aleph(x, e^{\ri \hat q})_{(x)} V(x,\hat p), \\
&& \vartheta (z(e^{\ri \hat q},  \hp), Z(e^{\ri \hat q},  \hp)) =
\aleph(x, e^{\ri \hat q})_{(x)}
\left(\eta(x, \hp) e^{\ri \hat q}\right)
\aleph(x, e^{\ri \hat q})_{(x)}^{-1},
\label{main}\eea
using (\ref{defoftx}) and (\ref{deff}).
The main point is that that this definition makes sense for all
$(e^{\ri \hat q}, \hat p)\in \bT(n) \times \bar\fC_x$.
One can also check by writing  explicit formulae that $\hat \pi_j$, $\cV$ and $\vartheta$ are
smooth functions on $\hat P_c$.

To present the explicit formulae of the above functions,
we first of all note that
\be
\hat \pi_1(z,Z)= - \log \vert Z\vert,
\quad
\hat\pi_k(z,Z)= -(k-1)x -\log \vert Z\vert - \sum_{j=1}^{k-1} \vert z_j\vert^2,
\quad k=2,\ldots, n.
\label{E1}\ee
Then we define the auxiliary functions
\be
Q_{j,k}:= \left[\frac{\hat\pi_j - \hat\pi_k - x}{\hat\pi_j - \hat\pi_k}\right]^{\frac{1}{2}},
\qquad
\forall  j\neq k \in\{1,\ldots, n\}.
\label{E2}\ee
As a result of the spectral gap condition (\ref{a7}), all expressions under the square root are non-negative,
and their denominators are non-zero. Of course, one can easily spell out the functions $Q_{j,k}$ more
explicitly. They depend only on the absolute values of the variables $Z$, $z_i$, and
satisfy, on the whole of $\hat P_c$, the inequalities
\be
Q_{j,k} >0
\quad\hbox{if}\quad   k-j\neq 1.
\label{E3}\ee
Then we have
\be
\cV_j = \frac{z_j}{\sqrt{\vert z_j\vert^2  + x}} \prod_{k\neq j,j+1} Q_{j,k},
\quad
j=1,\ldots, n-1
\quad\hbox{and}\quad
\cV_n =\prod_{k\neq n} Q_{n,k}.
\label{E4}\ee
The various entries of the matrix $\vartheta(z,Z)\in U(n)$ can be listed as follows:
\be
\vartheta_{i,i+1} = \frac{x}{\hat \pi_{i+1} - \hat \pi_i} \prod_{j\neq i,i+1} (Q_{i,j}\, Q_{j,i+1}),
\qquad
i=1,\ldots, n-1,
\label{E5}\ee
\be
\vartheta_{n,1}= \frac{x}{\hat \pi_1 - \hat \pi_n} \frac{\bar Z}{\vert Z\vert} \prod_{j\neq 1,n} (Q_{n,j}\, Q_{j,1}),
\label{E6}\ee
\be
\vartheta_{n,k}= \frac{x}{\hat \pi_k - \hat \pi_n}
\frac{\bar z_{k-1} Q_{n,k-1}}{\sqrt{\vert z_{k-1}\vert^2 +x}} \prod_{j\neq k-1,k,n} (Q_{n,j}\, Q_{j,k}),
\qquad k\in \{1,\ldots, n\}\setminus \{1,n-1,n\},
\label{E7}\ee
\be
\vartheta_{k,1}= \frac{x}{\hat \pi_1 - \hat \pi_k}
\frac{z_{k} Q_{k+1,1}}{\sqrt{\vert z_{k}\vert^2 +x}} \frac{\bar Z}{\vert Z\vert}
\prod_{j\neq 1,k,k+1} (Q_{k,j}\, Q_{j,1}),
\qquad k=2,\ldots, n-1,
\label{E8}\ee
\be
\vartheta_{k,k}=
\frac{z_{k} }{\sqrt{\vert z_{k}\vert^2 +x}}
\frac{\bar z_{k-1} }{\sqrt{\vert z_{k-1}\vert^2 +x}}
\Bigl(\prod_{j\neq k,k+1} Q_{k,j}\Bigr)
\Bigl(\prod_{l\neq k-1,k} Q_{l,k}\Bigr),
\qquad k=2,\ldots, n-1,
\label{E9}\ee
\be
\vartheta_{1,1}= \frac{z_1 \bar Z }{\vert Z\vert}
 \frac{\sqrt{ \vert z_{1}\vert^2 +2x}}{\vert z_{1}\vert^2 +x}\,
\prod_{j=3}^n (Q_{1,j} Q_{j,1}),
\quad
\vartheta_{n,n}= \bar z_{n-1}
 \frac{ \sqrt{\vert z_{n-1}\vert^2 +2x}}{\vert z_{n-1}\vert^2 +x}\,
\prod_{j=1}^{n-2} (Q_{n,j}\, Q_{j,n}),
\label{E10}\ee
and finally
\be
\vartheta_{a,b}=  \frac{z_a}{\sqrt{\vert z_{a}\vert^2 +x}}\frac{\bar z_{b-1}}{\sqrt{\vert z_{b-1}\vert^2 +x}}
\frac{x Q_{a,b-1}\, Q_{a+1,b}}{\hat\pi_b - \hat \pi_a}
\prod_{j\neq a,a+1, b-1, b} (Q_{a,j}\, Q_{j,b})
\label{E11}\ee
for the remaining indices $1\leq a, b\leq n$ subject to
$a\neq b$,
$b\neq a+1$,
$a\neq n$,
$b\neq 1$.

Now we are in the position to define the smooth map $\hat \cI: \hat P_c \to M_0$ by
\be
\hat \cI(z,Z):= ( \vartheta(z,Z)^{-1}, - \ri \hat\pi(z,Z), \xi(x, \cV(z,Z))).
\label{hatI}\ee
The rationale behind this definition, and also for the definition of the various
functions above, comes from the following relation:
\be
\hat \cI \circ \cZ_x = k_x.
\label{prop}\ee
Thus $\hat \cI$ is the unique continuous extension of the
map $k_x\circ \cZ_x^{-1}: \hat P_c^0 \to M_0$ to $\hat P_c$
(cf.~(\ref{3.69})).
Moreover, the map $\hat \cI$ enjoys the property
\be
\hat\cI(z(e^{\ri \hat q}, \hat p), Z(e^{\ri \hat q}, \hat p)) = K_x([\aleph(x, e^{\ri \hat q})_{(x)}], e^{\ri \hat q}, \hat p)
\qquad
\forall (e^{\ri \hat q}, \hat p)\in \bT(n) \times \bar\fC_x.
\label{propext}\ee
Recalling (\ref{kx}), note that (\ref{prop}) is  the restriction of ({\ref{propext})  to the
dense, open subset $\bT(n) \times \fC_x$.

\medskip
\noindent
{\bf Theorem 3.12.}
\emph{Consider the symplectic manifold $(\hat P_c, \hat \omega_c)$ (\ref{Pc}) and the  map $\hat \cI: \hat P_c \to
M_0$
defined by (\ref{hatI})  using the notations listed in (\ref{E1}-\ref{E11}).
Then $\hat \cI$ yields a smooth, global cross-section
 of the gauge orbits in the constraint-manifold
  $M_0= \Phi^{-1}(0) $ and it satisfies
\be
\hat \cI^* (\iota_0^* (\Omega_M)) = \hat \omega_c,
\label{pull}\ee
where $\iota_0: M_0 \to M$ is the embedding.
Therefore $(\hat P_c, \hat \omega_c)$ is a model of the reduced phase space.}

\medskip\noindent
{\bf Proof.}
The pull-back property (\ref{pull}) follows  by combining Proposition 3.11 with the property
$\cZ_x^*(\hat \omega_c) = \hat \omega$ (\ref{59}) and the fact that (\ref{prop}) holds.
In addition, we here use that $\cZ_x(\hat P) = \hat P_c^0$ is dense in $\hat P_c$.
We also conclude effortlessly, from (\ref{propext}) and Lemma 3.9, that the image of
$\hat \cI$ intersects every gauge orbit in $M_0$.

It only remains to show that the map $\hat \cI$ is injective and no two different points
on its image can be gauge equivalent.
This follows if we prove the following implication:
\be
A_{[h]}( \hat \cI(z,Z))= \hat\cI(z^\prime,Z^\prime)
\Longrightarrow h=\gamma \1_n\,(\gamma \in U(1)),\, z=z',\, Z=Z'.
\label{impl}\ee
By proving this statement, in which $h\in G$, we shall  also confirm again that $\bar G$ acts freely on $M_0$.
The proof of the above implication will rely on the following properties:

\begin{enumerate}

\item{The absolute values $\vert Z\vert$, $\vert z_i\vert$ are in one-to-one correspondence
with the values of the functions $\hat \pi_j$.}

\item{$\vartheta_{i,i+1}(z,Z)<0$ for all $i=1,\ldots, n-1$ and depends only on the absolute values $\vert Z\vert$,
$\vert z_j\vert$.}

\item{$\cV_n(z,Z)>0$ and depends only on the absolute values of $Z$ and the $z_j$.}

\item{$\cV_j(z,Z) = z_j f_j(z,Z)$ for $j=1,\ldots, n-1$, where $f_j(z,Z)>0$ and  $f_j$ depends only
 on the absolute values of $Z$ and the $z_k$.}

\item{$\vartheta_{n,1}(z,Z)= {\bar Z} f(z,Z)$, where $f(z,Z)>0$ and $f$ depends
only on the absolute values of $Z$ and the $z_k$.}

\end{enumerate}

Property 1 is clear from (\ref{E1}), and (\ref{E4}) implies properties 3 and 4.
Property 2 follows from (\ref{E3}) and property 5 from (\ref{E6}).

Now suppose that the equality
\be
A_{[h]}( \hat\cI(z,Z))= \hat\cI(z^\prime,Z^\prime)
\label{equal}\ee
holds and look at its `second component' in terms of writing the elements of $M$ as triples.
This gives
$
h \hat \pi(z,Z) h^{-1} = \hat \pi(z', Z').
$
Since the values of $\hat \pi$ are regular elements in the same Weyl alcove, we immediately
get that $h\in \bT(n)$ and $\hat \pi(z,Z)  =\hat \pi(z', Z')$.
By property 1, this proves that
$
\vert Z \vert = \vert Z'\vert
$,
and
$
\vert z_j \vert = \vert z_j'\vert$ for all $j=1,\ldots, n-1.
$
By looking at the $i, i+1$ component of the equality
$
h \vartheta(z,Z) h^{-1} = \vartheta(z',Z')
$,
which follows from (\ref{equal}), we get from property 2 and the previously established facts that
$h= \gamma \1_n$
for some $\gamma \in U(1)$.
Therefore we must have
$
\xi(x, \cV(z,Z)) = \xi(x, \cV(z', Z'))
$
and this implies now by property 3 that
$
\cV(z,Z) = \cV(z', Z').
$
Then property 4 entails that $z_j = z_j'$ for all $j$, and property 5 entails that $Z= Z'$.
\hspace*{\stretch{1}} \qedsymb

\medskip
\noindent
{\bf Definition 3.13.}
\emph{The Ruijsenaars dual  of the Sutherland system  of Definition 3.1} is
the integrable system given by
the commuting Hamiltonians  generated by the spectral invariants of the
unitary Lax matrix $\vartheta(z,Z)$ (\ref{main}) on the phase space $(\hat P_c, \hat \omega_c)$ (\ref{Pc}).
\medskip

Note that, because of Proposition 3.11, the system introduced above is the natural
completion of the
rational Ruijsenaars-Schneider system that possesses the Lax matrix $\hat L(\hat q, \hat p)$ (\ref{LRS}) on the
phase space $(\hat P,\hat \omega)$.
Here, $\hat P$ is identified with $\hat P_c^0$ (\ref{3.69}).
The same completion was introduced in \cite{RIMS95} without relying on
symplectic reduction.

\subsection{The duality map $\cR$}

Let us summarize the picture that emerges from the preceding two subsections.
First we recall from Section 2 that the symplectic reduction yields the
phase space $((T^*G)_\red, \Omega_\red)$ equipped with  two Abelian algebras
of integrable Hamiltonians, $\{ H_k\}$ and $\{ \hat H_{\pm k}\}$ (\ref{redHams}), whose
flows can be obtained as  projections of free flows.
This statement does not refer to any coordinate system or model:
the reduced phase space
$(T^*G)_\red$ is  the space of the gauge orbits $M_0/\bar G$.

In Section  3.1 we constructed the model $(P,\omega)\equiv (T^*Q(n), \Omega_{T^* Q(n)})$
 of the reduced phase space.
 When regarded as functions on $P$, the family $\{ H_k\}$ gives the commuting Sutherland
 Hamiltonians and the
family $\{ \hat H_{\pm k}\}$ gives the functions of the Sutherland position-variables
that represent
$S(n)$-invariant trigonometric polynomials on $\bT(n)^0$ (recall that $Q(n)=\bT(n)^0/S(n)$).
These latter functions separate the points of $Q(n)$, and fully determine the position data.

In Section 3.2 we constructed the model $(\hat P_c, \hat \omega_c)$ of the reduced phase space,
and have taken the reduced Hamiltonians $\{ \hat H_{\pm k}\}$ as the commuting Hamiltonians of
an integrable system, called the Ruijsenaars dual of the Sutherland system.
The dual system is a completion of the rational Ruijsenaars-Schneider system characterized by
the Hamiltonian $\hat H_{\mathrm{RS}}$ (\ref{HRS}) that leaves on $\hat P$, which is equivalent
to the dense, open submanifold $\hat P_c^0\subset \hat P_c$ (\ref{3.69}).
When viewed as functions on $\hat P_c$, the family $\{ H_k\}$ becomes equivalent to
the global position variables $\hat \pi_j \in C^\infty(\hat P_c)$ ($j=1,\ldots, n$) of
the completed Ruijsenaars-Schneider system.

Our construction automatically yields a natural symplectomorphism
\be
\cR: P \to \hat P_c.
\label{cR}\ee
The map $\cR$ sends the `Sutherland representative' of a point of $(T^*G)_\red$ to its
`Ruijsenaars representative'. This map  operates  by  gauge transformations
since $\hat P_c$ is realized as a global gauge slice in $M_0$ and $P$ is realized as the base of a
sub-bundle of $M_0$ with finite structure group $S(n)$.
The functions $\hat \pi_j \circ \cR \in C^\infty(P)$ define action variables
for the Sutherland system.
Upon restriction to the open dense submanifold that corresponds to $M_0^0 \subset M_0$,
where the strict spectral gap condition holds, the functions $\hat q_j \circ \cZ_x^{-1} \circ \cR$
define canonical conjugates of the Sutherland actions.
Conversely, when transferred to functions on $\hat P_c$, the Sutherland coordinates $q_j$
can be viewed as actions of the dual system.
More precisely, it is the trigonometric symmetric polynomials of the functions
$q_j \circ \cR^{-1}$ that provide  globally well-defined action variables for the dual system,
because $Q(n) = \bT(n)^0/S(n)$.

The above mentioned  properties of the `duality map' between $P$ and $\hat P_c$
were   established originally
in the impressive paper \cite{RIMS95} on the basis
of  very laborious, direct arguments.
In fact, our geometrically constructed  map $\cR$ (\ref{cR}) is precisely the
action-angle map of \cite{RIMS95}.
This holds since our construction of the map $\cR$ relies on the diagonalization
of $J$ for the triples $(g,J,\xi)\in M_0$ in the same way as the
construction of \cite{RIMS95} relies on the diagonalization
of the Sutherland Lax matrix.
The link is made by means of the relation given by (\ref{K.25}) with (\ref{K.12}).
One first obtains the identity of our $\cR$ and the action-angle map
of \cite{RIMS95} on the dense open submanifolds of the two models of $(T^*G)_\red$
corresponding to $M_0^0$ (\ref{3.55}), and then it holds also globally because our embedding (\ref{3.69})
of $\hat P$ into $\hat P_c$ is the same as the one used in \cite{RIMS95}.
The main advantage of the group theoretic approach is that the symplectic property of $\cR$
is guaranteed automatically.
One can also obtain
the integration algorithms of \cite{RIMS95} from the projections of the
  free flows displayed in (\ref{2.9})-(\ref{2.11}).
   This is a routine matter, and
 we refrain from presenting the details.
 The projected free flows are  complete on $(T^*G)_\red$ as  a result of general principles.
 In the present case it is readily seen that the projections of the flows (\ref{2.9}) respect the
  strict spectral gap condition,
so $M_0^0/\bar G$ is invariant under these flows, but
the projections of the `dual free flows' (\ref{2.9}), (\ref{2.10}) are complete only on the full reduced phase
space $M_0/\bar G$.

\section{Coverings and dualities}
\setcounter{equation}{0}

So far we have given a geometric interpretation to the lowest arrow
in (\ref{big}).
The aim of this section is to expound the web of dualities and coverings
for the three different versions of the Sutherland system \cite{RIMS95} described in the Introduction
and in Appendix A. The final result is represented by the diagram (\ref{D}), which is
an elaboration of the diagram (\ref{big}).

\subsection{Discrete symmetries and coverings before  KKS reduction}

To implement the ideas outlined in the Introduction,
we now consider the phase spaces
\be
M_2:= T^*(\bR \times SU(n)) \times \cO =
T^*\bR \times T^* SU(n) \times \cO
\ee
and
\be
M_1:= T^*(U(1) \times SU(n)) \times \cO =
T^*U(1) \times T^* SU(n) \times \cO
\ee
together with $M = T^* U(n) \times \cO$ that we dealt with so far.
Analogously to the case of $T^*U(n)$, we adopt the parametrization
\be
T^* SU(n) \simeq SU(n) \times su(n) = \{ (\Gamma, \cJ)\},
\quad
\Omega_{T^* SU(n)} = -d  {\operatorname{tr}}(\cJ d\Gamma \Gamma^{-1} )
\ee
and also adopt the identifications $T^*\bR = \bR \times \bR$ and $T^* U(1) = U(1) \times \bR$.
Therefore,
the symplectic manifold $(M_2, \Omega_{M_2})$ is realized as
\bea
&&M_2 =\bR \times \bR \times SU(n) \times su(n) \times \cO = \{ (u_0, w_0, \Gamma, \cJ, \xi)\},\nonumber\\
&&\Omega_{M_2} = d w_0 \wedge du_0 -d {\operatorname{tr}} (\cJ d\Gamma \Gamma^{-1}) + \Omega_\cO(\xi),
\eea
and we also have
\bea
&&M_1 =U(1)\times \bR \times SU(n) \times su(n) \times \cO = \{ (\zeta_0, v_0, \Gamma, \cJ, \xi)\},\nonumber\\
&&\Omega_{M_1} = d v_0 \wedge \frac{ d \zeta_0}{\ri \zeta_0} -d {\operatorname{tr}} (\cJ d\Gamma \Gamma^{-1}) + \Omega_\cO(\xi),
\eea
similarly to
\bea
&&M= T^* U(n) \times \cO = U(n) \times u(n) \times \cO = \{ (g,J,\xi)\},\nonumber\\
&&\Omega_M = - d {\operatorname{tr}} (J dg g^{-1}) + \Omega_\cO(\xi).
\eea

There is a symplectic action of the Abelian group
of the integers, $\bZ$, on $M_2$ generated by the symplectic diffeomorphism
$\theta(1)$ that implements the action of $1\in \bZ$ as follows:
\be
\theta(1): (u_0, w_0, \Gamma, \cJ, \xi) \mapsto (u_0 - \frac{2\pi}{n}, w_0, e^{\ri \frac{2\pi}{n}} \Gamma, \cJ, \xi).
\ee
We also consider the action of the subgroup $n\bZ < \bZ$ generated by
\be
\theta(1)^n: (u_0, w_0, \Gamma, \cJ, \xi) \mapsto (u_0 - 2\pi, w_0,  \Gamma, \cJ, \xi).
\ee
Moreover, a symplectic action of  $\bZ_n = \bZ/n\bZ$
(realized as the multiplicative group of the $n$-th roots of unity)
 on $M_1$  is generated by the following action
of the primitive $n$-th root of unity:
\be
\alpha(e^{\ri \frac{2\pi}{n}}): (\zeta_0, v_0, \Gamma, \cJ, \xi) \mapsto
(e^{-\ri \frac{2\pi}{n}} \zeta_0, v_0, e^{\ri \frac{2\pi}{n}} \Gamma, \cJ, \xi).
\ee
These actions arise from the cotangent lifts of corresponding actions of
central subgroups of $G_2$ and $G_1$ (\ref{1.8}) provided by
\bea
&& \{ (- k \frac{2\pi}{n}, e^{\ri k\frac{2\pi}{n}}\1_n)
\in \bR \times SU(n)\,\vert\, k\in \bZ\} < G_2,\nonumber\\
&& \{ (e^{-\ri k \frac{2\pi}{n}}, e^{\ri k\frac{2\pi}{ n}}\1_n)
\in U(1) \times SU(n)\,\vert\, k=0,1,\ldots, n-1\} < G_1,
\eea
which are isomorphic to $\bZ$ and  $\bZ_n$,  respectively.
The factorizations by these actions yield
the identifications
\be
M_1\simeq M_2/n\bZ,
\quad
M \simeq M_2/\bZ \simeq (M_2/ n\bZ)/(\bZ /n\bZ)\simeq M_1/\bZ_n.
\label{4.10}\ee
The projections responsible for these identifications are given by the maps
\be
\psi_2: M_2 \to M_1,
\quad
\psi_1: M_1 \to M,
\quad
\psi:= \psi_1\circ \psi_2: M_2 \to M,
\label{4.11}\ee
which can be written in terms of the above-introduced parametrizations  as
\be
\psi_2: (u_0, w_0, \Gamma, \cJ, \xi) \mapsto (e^{\ri u_0}, w_0, \Gamma, \cJ,\xi),
\quad
\psi_1: (\zeta_0, v_0, \Gamma, \cJ, \xi) \mapsto (\zeta_0\Gamma, \cJ + \frac{\ri}{n} v_0 \1_n, \xi).
\ee
These are symplectic coverings, i.e.,
\be
\psi_2^* (\Omega_{M_1}) = \Omega_{M_2}
\quad\hbox{and}\quad
\psi_1^*( \Omega_M)= \Omega_{M_1}.
\ee
We use these maps to pull-back the commuting families $\{ \cH_k\} \subset C^\infty(M)$
(\ref{freeH})  and
$\{\hat \cH_{\pm k}\}\subset C^\infty(M)$  (\ref{freehatH}) to $M_1$ and
 $M_2$, and thereby define the respective
`free' Hamiltonians
\be
\cH_j^1 := \cH_j \circ \psi_1,
\,\,
\hat \cH^1_{\pm k}:= \hat \cH_{\pm k}\circ \psi_1
\quad\hbox{and}\quad
\cH_j^2 := \cH_j \circ \psi,
\,\,
\hat \cH^2_{\pm k}:= \hat \cH_{\pm k}\circ \psi.
\label{4.15}\ee
One sees from the preceding considerations that
the three phase spaces $M_2$, $M_1$, $M$, together with their families of `free' Hamiltonians, are related
by reductions under the respective discrete symmetries represented by the $n\bZ$-action and the $\bZ$-action
on $M_2$, and the $\bZ_n$-action on $M_1$.

\subsection{$\bar G$-symmetry and KKS reduction on three levels}

In Section 3, we used the action of $\bar G= U(n)/U(1)$ on $M$ and analyzed the corresponding
KKS type symplectic reduction.
Recall that  every $[y]\in \bar G$ can be represented  by some $y\in SU(n)$,
and the $\bar G$-action $A_{[y]}: M\to M$ and its moment map $\Phi: M\to su(n)\simeq \mathrm{Lie}(\bar G)^*$
read
\be
A_{[y]}(g,J,\xi) = (ygy^{-1}, y Jy^{-1}, y\xi y^{-1}),
\quad
\Phi(g,J,\xi)= J - g^{-1} J g + \xi.
\ee
We now lift these to corresponding $\bar G$-actions and moment maps on $M_2$ and on $M_1$, furnished
respectively by
\be
A_{[y]}^2(u_0, w_0, \Gamma,\cJ,\xi) = (u_0,w_0, y\Gamma y^{-1}, y \cJ y^{-1}, y\xi y^{-1}),
\quad
\Phi_2(u_0, w_0, \Gamma,\cJ,\xi) = \cJ - \Gamma^{-1} \cJ \Gamma + \xi
\ee
and
\be
A_{[y]}^1(\zeta_0, v_0, \Gamma,\cJ,\xi) = (\zeta_0,v_0, y\Gamma y^{-1}, y \cJ y^{-1}, y\xi y^{-1}),
\quad
\Phi_1(\zeta_0, v_0, \Gamma,\cJ,\xi) = \cJ - \Gamma^{-1} \cJ \Gamma + \xi.
\ee
The $\bar G$-actions are trivial on $T^*\bR$ and on $T^*U(1)$, and by setting the moment maps to zero
they define the reduced phase spaces
\be
(T^* G_2)_\red:= \Phi_2^{-1}(0)/ \bar G = T^*\bR \times (T^* SU(n))_\red,
\ee
\be
(T^* G_1)_\red:= \Phi_1^{-1}(0)/ \bar G = T^*U(1) \times (T^* SU(n))_\red,
\ee
where
\be
T^* SU(n)_\red := (T^* SU(n) \times \cO)//_0 \bar G
\label{4.21}\ee
denotes the reduced phase space that arises from the KKS reduction of $T^* SU(n) \times \cO$.

The discrete symmetries described in the previous subsection commute with the relevant $\bar G$-symmetries,
that is, we have
\be
A^2_{[y]} \circ \theta(1) = \theta(1) \circ A^2_{[y]},
\quad
A^1_{[y]} \circ \alpha(e^{\ri \frac{2\pi}{n}})  =\alpha(e^{\ri \frac{2\pi}{n}}) \circ A^1_{[y]},
\ee
\be
\Phi_2 \circ \theta(1) = \Phi_2,
\quad
\Phi_1 \circ \alpha(e^{\ri \frac{2\pi}{n}}) = \Phi_1.
\ee
Moreover, we also have
\be
\psi_2 \circ A^2_{[y]} = A^1_{[y]}\circ  \psi_2,
\quad
\psi_1 \circ A^1_{[y]} = A_{[y]}\circ \psi_1,
\quad
\psi \circ A^2_{[y]} = A_{[y]}\circ \psi.
\ee
The above relations allow us to conclude that
it does not matter whether one  first performs `discrete reduction' (by the $\bZ$, $n\bZ$ or $\bZ_n$-action) and then
KKS reduction (by the $\bar G$-action) or the other way round, the final result will be the same.
In other words, there arises a natural $\bZ$-action (and $n \bZ$-action) on the reduced phase space
$(T^* G_2)_\red$ and a natural $\bZ_n$-action on $(T^* G_1)_\red$,
 generated say by
\be
 \theta(1)_\red: (T^* G_2)_\red \to (T^* G_2)_\red
\quad\hbox{and}\quad
\alpha(e^{\ri \frac{2\pi}{n}})_\red: (T^* G_1)_\red \to (T^* G_1)_\red,
\label{4.24}\ee
which induce the identifications
\bea
&&{(T^*G_1)}_\red\simeq {(T^*G_2)}_\red/n\bZ,\nonumber\\
&&
(T^*G)_\red \simeq {(T^*G_2)}_\red/\bZ = ( {(T^*G_2)}_\red/ n\bZ)/(\bZ /n\bZ)= {(T^*G_1)}_\red/\bZ_n.
\eea
These identifications are the remnants of those in (\ref{4.10}) that survive the KKS reduction.
In analogy to (\ref{4.11}), the identifications are associated with certain projections
\be
\psi_2^\red: {(T^*G_2)}_\red \to {(T^*G_1)}_\red,
\quad
\psi_1^\red: {(T^*G_1)}_\red \to {(T^*G)}_\red,
\quad
\psi_\red:= \psi_1^\red\circ \psi_2^\red.
\label{4.26}\ee
The phase space $(T^*G_i)_\red$ ($i=1,2$) carries the two Abelian algebras of integrable Hamiltonians
$\{ H_k^i\}$ and $\{\hat H^i_{\pm k}\}$ that can be characterized equivalently either as the KKS reductions
of the respective free Hamiltonians in (\ref{4.15}) or by means of the relations
\be
H_j^1 := H_j \circ \psi_1^\red,
\,\,
\hat H^1_{\pm k}:= \hat H_{\pm k}\circ \psi_1^\red
\quad\hbox{and}\quad
H_j^2 := H_j \circ \psi_\red,
\,\,
\hat H^2_{\pm k}:= \hat H_{\pm k}\circ \psi_\red.
\label{4.15red}\ee
To describe the projections (\ref{4.26}) explicitly,
we need a concrete description of the `building block' $(T^*SU(n))_\red$ that
appears as a factor both in ${(T^*G_2)}_\red$ and in ${(T^*G_1)}_\red$.
Similarly to the $U(n)$-case, we  actually have two such descriptions.

\subsection{Two models of $(T^* SU(n))_\red$ in duality}

Referring to the notations of Appendix A.1,
consider the cotangent bundle
\be
T^*SQ(n) \simeq  T^* \operatorname{Simp}_{n-1} \simeq  \operatorname{Simp}_{n-1}\times \bR^{n-1} = \{(\delta, \gamma)\},
\quad
\Omega_{T^* SQ(n)} = \sum_{j=1}^{n-1} d \gamma_j \wedge d\delta_j.
\ee
By using $\beta_k(\delta)$ in (\ref{A15}) and
$\beta(\delta):= \diag(\beta_1(\delta),\ldots, \beta_n(\delta))$
for $\delta \in \operatorname{Simp}_{n-1}$ (\ref{A5}),
introduce the $su(n)$-valued function
\be
\cJ(\delta,\gamma):= \ri \sum_{j=1}^{n-1} \gamma_j (E_{j,j} - E_{j+1, j+1}) + \ri x\sum_{a\neq b}
\frac{E_{a,b}}{ 1 - e^{\ri (\beta_b(\delta) - \beta_a(\delta))}}.
\ee
By setting $\xi_0:= -\ri x \sum_{a\neq b} E_{a,b}$, the manifold
\be
(T^* SU(n))_\red^{\mI}:= \{ (e^{\ri \beta(\delta)}, \cJ(\delta,\gamma), \xi_0)\,\vert\, (\delta, \gamma)\in
T^*\operatorname{Simp}_{n-1}\} \simeq T^* SQ(n)
\label{4.30}\ee
is a model of  the reduced phase space defined by (\ref{4.21}).
In fact,  $(T^* SU(n))_\red^\mI\subset T^*SU(n) \times \cO $ is a \emph{global cross-section} for the action of $\bar G$ on the
zero level set of the moment map $\Phi_0: T^* SU(n) \times \cO \to su(n)$ given by
\be
\Phi_0( \Gamma, \cJ, \xi) = \cJ - \Gamma^{-1} \cJ \Gamma + \xi,
\ee
and the pull-back of the symplectic form of $T^* SU(n) \times \cO$ on $(T^* SU(n))_\red^\mI$ coincides with
$\Omega_{T^* SQ(n)}$.
The proof of this result \cite{KKS} follows the lines of Section 3.1.
The model (\ref{4.30}) of $(T^*SU(n))_\red$ is naturally associated with the `relative motion'
(i.e., motion in  the center of mass frame)  of
the $n$ distinguished particles on the circle; see also Appendix A.
The relative motion is governed by the
 Hamiltonian $-\frac{1}{2} {\operatorname{tr}}(\cJ(\delta, \gamma)^2)$,
which is Liouville integrable on account of the commuting family given by the spectral invariants
of $\cJ(\delta, \gamma)$.

Our next goal is to identify a dual model of the reduced phase space
$(T^* SU(n))_\red$ with
\be
 \bC^{n-1} = \{ \zeta= (\zeta_1,\ldots, \zeta_{n-1}) \},
\quad
\Omega_{\bC^{n-1}} := \ri \sum_{j=1}^{n-1} d \zeta_j \wedge d \bar \zeta_j,
\label{4.33}\ee
which can be recast equivalently as the standard symplectic vector space of $\bR^{2(n-1)}$.
For this purpose, by a suitable modification of the formulas of Section 3.2, we construct
a second global cross-section in $\Phi_0^{-1}(0) \subset T^* SU(n) \times \cO$
as follows.
First, we define the functions
\be
\hat \pi_k^0(\zeta):= x\frac{n+1-2k}{2} - \sum_{1\leq j\leq (k-1)}\frac{j}{n} \vert \zeta_j\vert^2
+ \sum_{k\leq j\leq (n-1)}\frac{n-j}{n} \vert \zeta_j\vert^2,
\qquad
\forall k=1,\ldots, n,
\ee
so that we have $\sum_{k=1}^n \hat\pi_k^0(\zeta)=0$.
Second, we introduce the auxiliary functions $Q_{j,k}^0(\zeta)$ by replacing $\hat \pi$ in (\ref{E2}) with $\hat \pi^0$.
Third, we define the $SU(n)$-valued function $\vartheta^0(\zeta)$ by means of the replacements
\be
Z\to 1,\quad
z_j \to \zeta_j,
\quad
\hat \pi_j \to \hat\pi_j^0,
\quad
Q_{j,k} \to Q_{j,k}^0
\ee
in the formulas (\ref{E5})-(\ref{E11}), and analogously define the $\bC^n$-valued function
$\cV^0(\zeta)$ by modifying (\ref{E4}).
Finally, similarly to the formula (\ref{hatI}), we define the smooth map
\be
\hat \cI^0: \bC^{n-1} \to \Phi_0^{-1}(0)\subset T^*SU(n) \times \cO,
\qquad
\hat \cI^0(\zeta):= \bigl(\vartheta^0(\zeta)^{-1}, -\ri \hat\pi^0(\zeta),  \xi(x, \cV^0(\zeta))\bigr).
\ee
It can be shown that $\hat \cI^0$ is injective and  its image,
\be
(T^*SU(n))_\red^{\mII} := \{ (\vartheta^0(\zeta)^{-1}, -\ri \hat\pi^0(\zeta),  \xi(x, \cV^0(\zeta)))\, \vert\,
\zeta \in \bC^{n-1}\} \simeq \bC^{n-1},
\ee
is a \emph{global cross-section of the $\bar G$-orbits that converts the reduced symplectic form into
$\Omega_{\bC^{n-1}}$}.
Of course, this model of the reduced phase space carries the  distinguished commuting
Hamiltonians provided by the spectral invariants of $\vartheta^0(\zeta)$.

The above construction yields automatically the `duality symplectomorphism in the center of mass frame'
\be
\cR_0:(T^* SU(n))_\red^{\mI}  \to (T^* SU(n))_\red^{\mII},
\label{4.35}\ee
which operates by the pertinent gauge transformations between the two global cross-sections.
The map $\cR_0$  can be viewed as an action-angle transform for the center of mass version of the Sutherland system,
and its inverse is  an action-angle transform for the $SU(n)$-version of the completed
dual Ruijsenaars-Schneider system.

\subsection{The web of discrete reductions and dualities}

The preceding constructions imply
that the dual phase spaces associated with $U(n)$,
\be
P \simeq (T^*G)_\red \simeq \hat P_c,
\label{4.38}\ee
possess the symplectic covering spaces given by the dual pairs
\be
T^* U(1) \times (T^*SU(n))_\red^\mI \simeq (T^* G_1)_\red \simeq  T^* U(1) \times (T^*SU(n))_{\red}^{\mII}
\label{4.39}\ee
and
\be
T^* \bR \times (T^*SU(n))_\red^\mI \simeq (T^* G_2)_\red \simeq  T^* \bR \times (T^*SU(n))^{\mII}_\red.
\label{4.40}\ee
The respective dual pairs appear on the two ends of the three  chains of symplectomorphisms
in (\ref{4.38})-(\ref{4.40}), while the middle
term (such as $(T^*G)_\red$ etc)
  refers to the `abstract reduced phase space' that exists as a space of orbits, irrespective
of any model of it.
The dual systems of integrable Hamiltonians are provided by
  $\{ H_k^i\}$ and $\{\hat H^i_{\pm k}\}$  (\ref{4.15red}) expressed in terms of the alternative models of $(T^* G_i)_\red$
 in analogy to the Hamiltonians $\{H_k\}$ and $\{ \hat H_{\pm k}\}$ on $(T^*G)_\red$.
In more detail, the situation is depicted by the following commutative diagram:
\be
\xymatrix{
T^*{\mathbb R}\times T^* SQ(n)
\ar[rr]^-{\mathrm{id}_2 \times {\mathcal R}_0}
\ar[d]_-{\psi_2^\mI} &&
T^*{\mathbb R}\times {\mathbb C}^{n-1} \ar[d]^-{\psi_2^{\mII}}\\
T^*U(1) \times T^* SQ(n)   \ar[rr]^-{\mathrm{id}_1 \times {\mathcal R}_0}
\ar[d]_-{\psi_1^\mI} &&
T^*U(1) \times  {\mathbb C}^{n-1} \ar[d]^-{\psi_1^{\mII}}\\
P =T^*Q(n)\ar[rr]^-{\mathcal R}&&
{\hat P}_c=  {\mathbb C}^{n-1} \times {\mathbb C}^\times }
\label{D}\ee
This diagram is the detailed version of the diagram (\ref{big}) presented the Introduction,
where now we have a rather complete understanding of all its ingredients.
In particular, $\psi_2^{\mI}$  in (\ref{D}) represents the projection $\psi_2^\red$ (\ref{4.26})
of  the $n\bZ$-reduction in terms of the models
\bea
&&(T^* G_2)_\red \simeq T^*\bR \times (T^* SU(n))_\red^\mI \equiv T^*\bR \times T^* SQ(n),\nonumber\\
&&(T^* G_1)_\red \simeq T^*U(1) \times (T^* SU(n))_\red^\mI \equiv T^*U(1) \times T^* SQ(n),
\eea
and $\psi_2^{\mII}$ represent $\psi_2^\red$  in terms of the models
\bea
&&(T^* G_2)_\red \simeq T^*\bR \times (T^* SU(n))_\red^\mII \equiv T^*\bR \times \bC^{n-1},\nonumber\\
&&(T^* G_1)_\red \simeq T^*U(1) \times (T^*SU(n))_\red^\mII \equiv T^*U(1) \times \bC^{n-1}.
\eea
The map $\cR_0$ is the `duality symplectomorphism in the center of mass frame' (\ref{4.35}), while
$\mathrm{id}_2$ and $\mathrm{id}_1$ denote the identity maps of $T^*\bR$ and $T^* U(1)$, respectively.
The maps $\psi_2^\mI$ and $\psi_2^\mII$ are quite simple, since they have factorized form and their non-trivial
factor is the obvious map $T^* \bR \to T^* U(1)$ that sends $(u,w)$ to $(\zeta_0,v_0):=(e^{\ri u}, w)$.
The maps $\psi_1^\mI$ and $\psi_1^\mII$ are
the projections induced by the $\bZ_n$-action,
$\alpha(e^{\ri \frac{2\pi}{n}})_\red: (T^* G_1)_\red \to (T^* G_1)_\red$, that admits
the alternative realizations
\bea
&&\alpha(e^{\ri \frac{2\pi}{n}})_\red^\mI: T^*U(1) \times T^* SQ(n) \to T^*U(1) \times T^* SQ(n), \\
&&\alpha(e^{\ri \frac{2\pi}{n}})_\red^\mI:(\zeta_0, v_0, \delta, \gamma) \mapsto
(e^{-\ri \frac{2\pi}{n}} \zeta_0, v_0, \delta_2, \ldots, \delta_{n-1},
2\pi - \sum_{j=1}^{n-1} \delta_j, \gamma_2 - \gamma_1, \ldots, \gamma_{n-1} - \gamma_1, -\gamma_1),\nonumber
\eea
and
\bea
&&\alpha(e^{\ri \frac{2\pi}{n}})_\red^\mII: T^*U(1) \times \bC^{n-1} \to T^*U(1) \times \bC^{n-1},\\
&&\alpha(e^{\ri \frac{2\pi}{n}})_\red^\mII: \bigl(\zeta_0,v_0, \{\zeta_j\}_{j=1}^{n-1} \bigr)
\mapsto \bigl(e^{- \ri \frac{2\pi}{n}}\zeta_0,v_0,
\{ e^{\ri \frac{2\pi}{n}(n-j)}\zeta_j\}_{j=1}^{n-1}\bigr).
\nonumber
\eea
These formulas can be obtained directly from the definitions by `diagram chasing', and agree with
corresponding formulae in \cite{RIMS95}. The associated formula of $\psi_1^\mII$ is found to be
\be
\psi_1^\mII: (\zeta_0, v_0, \zeta)\mapsto (z,Z)
\,\,\,\hbox{with}\,\,\,
z_j = \zeta_0^{n-j} \zeta_j,
\,\,
Z=
\zeta_0^n\exp\Bigl(\frac{1-n}{2} x  + \frac{v_0}{n} +\sum_{k=1}^{n-1} \frac{k-n}{n} \vert \zeta_k \vert^2  \Bigr),
\ee
and   $\psi_1^\mI$ is described in Appendix A.3.

We
constructed the duality map $\cR$ in Section 3, and now we observe that $\cR$ can be characterized
as the unique map that makes the diagram (\ref{D}) commute.
All the maps that appear in (\ref{D}) can be found\footnote{In continuation of
the footnote given after (\ref{big}), we remark that the symbols $T^*SQ(n)$, $\bC^{n-1}$ and $\cR_0$  in (\ref{D})
correspond respectively to the symbols $M$, $\hat M$ and $\phi$ in (1.74) of \cite{RIMS95}.
Note also that in \cite{RIMS95}  the transposition
of the two components of $\bR^2$
is used in  place of our $\mathrm{id}_2$ in (\ref{D}) due to an immaterial difference of conventions.}
in the paper \cite{RIMS95}, too, and
there the investigation  of the properties of $\cR$ was actually based on a corresponding commutative diagram.
Our geometric derivation reveals  the natural group theoretic origin of the web of dualities and coverings
encoded by (\ref{D}).

\section{Discussion}
\setcounter{equation}{0}

In this paper we presented a group theoretic interpretation of the duality relation
between the trigonometric Sutherland system and the completion of a certain real form of the complex rational
Ruijsenaars-Schneider system.
More precisely, we dealt with three
variants of these dual pairs and connected their covering Poisson maps to
the covering homomorphisms in (\ref{1.8}).
All ingredients of our diagram (\ref{D}) were constructed previously by Ruijsenaars \cite{RIMS95}
relying on direct methods, i.e., without using symplectic reduction.
The powerful tool of symplectic reduction
allowed us to shed a new light on the  web of dualities and coverings, and it also allowed us to
simplify the original arguments of \cite{RIMS95}. In particular, the symplectic character of the
duality maps is obvious in our setting, while originally this required a complicated proof.

As was briefly mentioned also in the Introduction,
three other cases of Ruijsenaars' duality relations \cite{SR-CMP,RIMS95} were treated before
by reductions of finite-dimensional, real symplectic manifolds.
First, one can  explain
the self-duality of the rational Calogero system with the help of the classical KKS reduction \cite{KKS}
of $T^*u(n)$.
Second,  one can obtain \cite{FK-JPA}
 the duality between the standard hyperbolic Sutherland and  rational Ruijsenaars-Schneider systems
 by reduction of the phase space
$T^* \cP(n)$,
where $\cP(n)$ is the symmetric space of positive definite Hermitian matrices.
Third, the dual pair involving the standard trigonometric Ruijsenaars-Schneider system can be interpreted
in terms of a  Poisson-Lie analogue of the KKS reduction \cite{FK3}.
In the last case the unreduced
phase space is the Heisenberg double of the Poisson-Lie group $U(n)$. This Poisson-Lie
analogue of $T^*U(n)$ is the real Lie group $GL(n,\bC)$ equipped with a certain
symplectic structure.
The `relativistic' generalization \cite{RIMS95} of the diagram (\ref{D}) can also be
obtained by reducing covering spaces of $GL(n,\bC)$.

However,   at the moment of writing,  there still exist two cases of the duality established in \cite{RIMS95}
for which an interpretation of the above type is not known.
These concern the self-dual systems provided on the one hand by the standard,  physically most important,
hyperbolic Ruijsenaars-Schneider system, and on the other by the
so-called $\mathrm{III}_{\mathrm b}$ real form of the complex trigonometric
Ruijsenaars-Schneider system \cite{RIMS95}.
Motivated by the gross features of these systems, we find it tempting to speculate that they should be the
reductions of suitable $U(n)$-symmetric `free systems' on manifolds of the form
\be
\cP(n) \times \cP(n)
\quad
\hbox{and}\quad
U(n) \times U(n).
\ee
Specifically, we expect that the $\mathrm{III}_{\mathrm b}$  real form
(also known as  the compactified trigonometric system \cite{VV})
can be derived by applying q-Hamiltonian reduction to $U(n)\times U(n)$ equipped with
the structure of the fused double defined in  \cite{AMM}, and this
will yield the correct finite-dimensional counterpart of the infinite-dimensional reduction
suggested in \cite{GN}. This issue is currently under investigation.
It is a very intriguing question  whether it is possible to construct
a suitable  $U(n)$-symmetric  `free system'  on
$\cP(n) \times \cP(n)$ so that it could serve as
the starting point for  the derivation of the hyperbolic Ruijsenaars-Schneider system by  reduction.
We stress that these questions concern systems with \emph{real} particle-positions, and hence their solution
requires to go beyond the treatment of the \emph{complex} trigonometric system
presented in \cite{FR,Obl}.

It is also an open problem to extend the Ruijsenaars dualities to systems with two types of particles as well
as to $BC(n)$ systems.
In the former case the action-angle maps, without an interpretation in terms of dualities,
were described in \cite{RIMS94}.
In our opinion these problems are important and non-trivial.
They could pose a worthwhile challenge for the interested reader.

\renewcommand{\theequation}{\arabic{section}.\arabic{equation}}
\renewcommand{\thesection}{\Alph{section}}
\setcounter{section}{0}

\section{Three variants of the Sutherland phase space}
\setcounter{equation}{0}
\renewcommand{\theequation}{A.\arabic{equation}}

In this appendix we explain how the phase spaces $P_2$, $P_1$ and $P$ displayed in (\ref{1.4})
correspond, respectively,  to distinguishable particles
moving  on the line $\bR$ or on the circle $U(1)$, or to
indistinguishable particles on the circle.
Our treatment here is  close to \cite{RIMS95}, but we pay
more attention to the Lie-theoretic interpretation of the pertinent
configuration spaces.

\subsection{Distinguishable particles on the line}

Consider $n$ distinguishable particles on the line interacting according to the Hamiltonian of the form
(\ref{1.1}).
Due to the repulsive periodic potential, the order of the particles cannot change during the motion
and the distance between the particles is also bounded. One possible choice from equivalent
configuration spaces is thus the convex domain
\be
C(n):= \{ u\in \bR^n\,\vert \, u_1 > u_2 > \cdots >u_n,\,\, u_1 - u_n < 2\pi\}.
\label{A1}\ee
The corresponding phase space,
\be
T^* C(n) = C(n) \times \bR^n = \{ (u, w)\,\vert\, u\in C(n),\, w\in \bR^n\},
\quad
\Omega_{T^* C(n)} = \sum_{j=1}^n dw_j \wedge d u_j,
\label{A2}\ee
is equipped with the Hamiltonian
\be
H^{T^* C(n)}_{\mathrm{Suth}}(u,w) = \frac{1}{2}\sum_{j=1}^n w_j^2 +
\frac{1}{4} \sum_{1\leq i<j\leq n} \frac{x^2}{\sin^2\!\left(\frac{u_i - u_j}{2}\right)}.
\label{A3}\ee
The configuration space $C(n)$ can be represented as the Cartesian product,
\be
C(n) \simeq \bR \times \operatorname{Simp}_{n-1},
\label{A4}\ee
where the line $\bR$ belongs to the center of mass motion,
and the open simplex,
\be
\operatorname{Simp}_{n-1}:= \{ \delta \in \bR^{n-1}\,\vert\,
\delta_j>0, \quad
\sum_{j=1}^{n-1} \delta_j < 2\pi\},
\label{A5}\ee
is the configuration space of the relative motion.
We denote the center of mass coordinate by $u_0$, its canonical conjugate by $w_0$, and the canonical
conjugates of the relative coordinates $\delta_j$ by $\gamma_j$.
With these notations, the `separated form' of the phase space is
\be
T^* C(n) = T^* \bR \times T^* \operatorname{Simp}_{n-1} \equiv (\bR\times \bR) \times
(\operatorname{Simp}_{n-1} \times \bR^{n-1})= \{ (u_0, w_0)\} \times \{ (\delta, \gamma)\}
\label{A6}\ee
with the symplectic form
\be
\Omega_{T^* C(n)}=\Omega_{T^* \bR} + \Omega_{T^* \operatorname{Simp}_{n-1}} = dw_0 \wedge du_0 +
\sum_{j=1}^{n-1} d \gamma_j \wedge d \delta_j.
\label{A7}\ee
The map between the above two systems of Darboux coordinates on $T^* C(n)$ is provided by the following formulae:
\bea
&& \delta_j = u_j - u_{j+1},\qquad
\gamma_j = \sum_{k=1}^j w_k - \frac{j}{n} \sum_{k=1}^n w_k,
\quad j=1,\ldots, n-1, \nonumber\\
&& u_0 = \frac{1}{n} \sum_{k=1}^n u_k,\qquad
w_0 = \sum_{k=1}^n w_k.
\label{A8}\eea
The inverse formulae are
\bea
&& u_n=u_0 - \frac{1}{n} \sum_{k=1}^{n-1} k \delta_k, \quad
u_j= u_0 - \frac{1}{n} \sum_{k=1}^{n-1} k \delta_k + \sum_{k=j}^{n-1} \delta_k,
\quad  j=1,\ldots,n-1, \nonumber\\
&& w_m= \gamma_m - \gamma_{m-1} + \frac{1}{n} w_0,
\quad
m=1,\ldots, n, \quad
(\gamma_0= \gamma_n := 0).
\label{A9}\eea
When expressed in terms of the `separated variables', the kinetic energy becomes
\be
\frac{1}{2}\sum_{k=1}^n w_k ^2 = \frac{1}{2n} w_0^2 + \frac{1}{2}\sum_{j,k=1}^{n-1} A_{j,k} \gamma_j \gamma_k,
\label{A10}\ee
where
\be
A_{j,k} := {\operatorname{tr}}\left((E_{j,j} - E_{j+1,j+1}) ( E_{k,k} - E_{k+1,k+1})\right)
\label{A11}\ee
is the Cartan matrix of $sl(n)$; the potential energy depends only on the relative coordinates
$\delta_j$.

For our purposes, it is important to note that $\operatorname{Simp}_{n-1}$ can be regarded as a model of
the group theoretically natural configuration space
\be
SQ(n):= S\bT(n)^0/S(n),
\label{A12}\ee
which is the space of Weyl orbits in the regular part of the maximal torus $S\bT(n) < SU(n)$.
To explain this, introduce the open Weyl alcove
\be
\cA(n-1):= \{ \beta=\diag(\beta_1, \beta_2, \ldots, \beta_n)\,\vert\, \beta_1 > \beta_2 > \cdots > \beta_n,\,\,
\beta_1 -\beta_n < 2\pi,\,\, {\operatorname{tr}}(\beta)=0\}.
\label{A13}\ee
The exponential map can be used to map $\cA(n-1)$ diffeomorphically onto the open submanifold
\be
A(n-1):=\{e^{\ri \beta}\,\vert\, \beta \in \cA(n-1)\} \subset ST(n)^0,
\label{A14}\ee
which is a fundamental domain for the $S(n)$-action on $S\bT(n)^0$.
Moreover,  we define a diffeomorphism between $\cA(n-1)$ and the simplex
$\operatorname{Simp}_{n-1}$ by the map
$\delta \mapsto \beta(\delta)$ given by
\be
\beta_n(\delta) = -\frac{1}{n} \sum_{k=1}^{n-1} k \delta_k,
\qquad
\beta_j (\delta) = \beta_n(\delta) + \sum_{k=j}^{n-1}  \delta_k,
\quad j=1,\ldots, n-1.
\label{A15}\ee
Altogether we have the identifications
\be
SQ(n) \longleftrightarrow A(n-1) \longleftrightarrow \cA(n-1)
\longleftrightarrow \operatorname{Simp}_{n-1}.
\label{A16}\ee

\subsection{Distinguishable particles on the circle}

The manifold $\bT(n)^0$ (\ref{K.1}) can be viewed as the set of possible configurations of $n$
distinguishable `non-coinciding point particles' moving on the unit circle $U(1)$.
We attach the labels $1,\ldots, n$ to the distinguished particles and identify
$\tau= \diag(\tau_1,\ldots, \tau_n) \in \bT(n)^0$
as the configuration for which $\tau_k$ is the location of the particle with label $k$.
It is easy to see that the connected components of the manifold $\bT(n)^0$ correspond
to the different possible cyclic orderings
of the $n$ distinct particles.
The cyclic orderings correspond, in turn, to the $(n-1)!$
different $n$-cycles in the group $S(n)$.
We can  restrict the dynamics to a single connected component, and we choose the particular one
given by the manifold
\be
K(n):= \{ e^{\ri q}\,\vert\, q=\diag(q_1, \ldots, q_n),\,\, q_1 > q_2 > \cdots > q_n,
\,\, q_1 - q_n < 2\pi\}.
\label{A18}\ee
That is, we take the phase space of the distinguished particles to be the symplectic manifold
\be
T^* K(n) = K(n) \times \bR^n = \{ (e^{\ri q}, p)\},
\quad
\Omega_{T^* K(n)} = \sum_{k=1}^n d p_k \wedge \frac{d e^{\ri q_k}}{ \ri e^{\ri q_k}}.
\label{A19}\ee
It should be noted that the variable $q_n$ is ambiguous up to multiples of $2\pi$,
but $e^{\ri q_n}$ is well-defined
and once $q_n$ is chosen then the other $q_j$ are uniquely determined by the
conditions specified in (\ref{A18}).
Taking this into account, globally well-defined smooth coordinates on $K(n)$ are provided by
\be
\zeta_0:= e^{ \ri q_n} \exp({\frac{\ri}{n} \sum_{j=1}^{n-1} j (q_j - q_{j+1})}) \quad
\hbox{and}\quad
\delta_j := q_j - q_{j+1},
\quad j=1,\ldots, n-1.
\label{A20}\ee
These formulae define the map $K(n) \to  U(1) \times \operatorname{Simp}_{n-1}$,
\be
e^{\ri q} \mapsto (\zeta_0(e^{\ri q}), \delta(e^{\ri q})).
\label{A21}\ee
This is  a diffeomorphism with the inverse map $U(1) \times \operatorname{Simp}_{n-1} \to K(n)$ furnished by
\be
(\zeta_0, \delta) \mapsto \zeta_0 e^{\ri \beta(\delta)},
\label{A22}\ee
where $\beta_k(\delta)$ is given by (\ref{A15}).
As a consequence, we obtain the identification
\be
T^* K(n) = T^* U(1) \times T^* \operatorname{Simp}_{n-1} \equiv (U(1) \times \bR) \times
(\operatorname{Simp}_{n-1} \times \bR^{n-1}) = \{ (\zeta_0, v_0)\} \times \{(\delta, \gamma)\},
\label{A23}\ee
whereby we can write the symplectic form as
\be
\Omega_{T^*K(n)}\simeq \Omega_{T^* U(1)} + \Omega_{T^*\operatorname{Simp}_{n-1}} = d v_0 \wedge \frac{d \zeta_0}{\ri \zeta_0}
+ \sum_{j=1}^{n-1} d \gamma_j \wedge d \delta_j.
\label{A24}\ee
The canonical momenta $v_0$ and $\gamma_j$ are related to the momenta $p_k$ of the individual
particles according to
\be
v_0 = \sum_{k=1}^n p_k,
\quad
\gamma_j = \sum_{k=1}^j p_k - \frac{j}{n} \sum_{k=1}^n p_k.
\label{A25}\ee
By using the pertinent identification in (\ref{A16}), the second term in
(\ref{A24}) can be regarded also as the symplectic form of $T^* SQ(n)\simeq T^*\operatorname{Simp}_{n-1}$.

Now some remarks are in order.
First, note that $\zeta_0$ (\ref{A20}) defines a `center of mass' for the distinguishable particles
moving on the circle. In fact, $\zeta_0$  gets rotated by the angle $\alpha$ if all the particle positions
are  rigidly rotated by the same angle $\alpha$;  and
 the relative coordinates $\delta_j$ do not change under these rigid rotations.

Second, let us notice that the phase space of the particles on the circle is actually
a symplectic quotient of the phase space of the particles on the line.
Indeed, there is a free, properly discontinuous, symplectic  action of the group $n\bZ<\bZ$ on
$T^*C(n)$ generated by the action of $n\in n\bZ$ defined by
$
 ((u_1,\ldots, u_n), w) \mapsto ((u_1-2\pi,\ldots, u_n-2\pi), w)
$,
which translates all particle coordinates by $2\pi$.
Equivalently, it acts on the separated variables in (\ref{A6}) according to
$
  (u_0, w_0, \delta, \gamma) \mapsto (u_0 - 2\pi, w_0,\delta,\gamma)
$.
The corresponding space of orbits is naturally a symplectic manifold,
and we can make the identification
\be
T^* C(n)/ n\bZ \equiv T^* K(n).
\label{A28}\ee
The associated projection
\be
\psi_2^{\mI}: T^* C(n) \to T^* K(n)
\label{A29}\ee
is given explicitly as
\be
(u, w) \mapsto (e^{\ri q}, p):= (e^{\ri\, \diag(u_1,\ldots, u_n)}, w),
\label{A30}\ee
or in terms of the separated variables displayed in (\ref{A6}) and (\ref{A23}) simply as
\be
 T^*\bR \times T^*\operatorname{Simp}_{n-1} \ni(u_0, w_0, \delta, \gamma)
 \mapsto (e^{\ri u_0}, w_0, \delta, \gamma) \in T^*U(1) \times T^*\operatorname{Simp}_{n-1}.
\label{A31}\ee
Thus the only effect of the factorization  by the $n\bZ$-action
is to identify the variable
$\zeta_0\in U(1)$ as the exponential of the center of mass $u_0$ of the particles
on the line.
The projection $\psi_2^\mI$ is locally symplectic, and   $T^*C(n)$ is a symplectic $n\bZ$-covering of $T^* K(n)$.
On account of (\ref{A16}), the notation $\psi_2^\mI$ (\ref{A29}) is consistent with the diagram (\ref{D}).

\subsection{Indistinguishable particles on the circle}

The permutation group $S(n)$ acts freely on $\bT(n)^0$ by the formula
\be
\sigma(\tau)_k:= \tau_{\sigma^{-1}(k)},
\quad
\forall \sigma\in S(n),\,\, \forall \tau=\diag(\tau_1,\ldots, \tau_n)\in \bT(n)^0.
\label{A32}\ee
By definition, the configuration space $Q(n)$ is obtained from $\bT(n)^0$ by identifying
the elements that are related by permutations. Thus $Q(n)$ may describe indistinguishable particles,
or distinct particles whose distinction is erased when recording the configurations.
It is clear that every element of $\bT(n)^0$ can brought into the connected component
$K(n)\subset \bT(n)^0$ (\ref{A18})  by a suitable
permutation, and the subgroup of $S(n)$ that maps $K(n)$ to $K(n)$ is generated by the cyclic permutation,
$\mu \in S(n)$ given by $\mu: (1,2, \ldots, n)\mapsto (\mu(1), \mu(2), \ldots, \mu(n)):=(n,1,\ldots, n-1)$.
Denoting this subgroup as $\bZ_n < S(n)$, we obtain the identification
\be
Q(n)= \bT(n)^0/ S(n) = K(n)/\bZ_n.
\label{A33}\ee
In terms of the model $K(n) \equiv U(1) \times \operatorname{Simp}_{n-1}$ given by (\ref{A21}), the
cyclic permutation acts as
\be
\mu: (\zeta_0, \delta_1,\ldots, \delta_{n-2},\delta_{n-1})
\mapsto (e^{-\ri \frac{2\pi}{n}} \zeta_0, \delta_2, \ldots, \delta_{n-1},
2\pi - \sum_{j=1}^{n-1} \delta_j),
\label{A34}\ee
and the cotangent lift of this  action reads
\be
\mu:  (\zeta_0, v_0, \delta, \gamma) \mapsto
(e^{-\ri \frac{2\pi}{n}} \zeta_0, v_0, \delta_2, \ldots, \delta_{n-1},
2\pi - \sum_{j=1}^{n-1} \delta_j, \gamma_2 - \gamma_1, \ldots, \gamma_{n-1} - \gamma_1, -\gamma_1).
\label{A35}\ee
If, by using that $\operatorname{Simp}_{n-1}\simeq SQ(n)$ according to (\ref{A16}),
 we identify $K(n)$ with $U(1) \times SQ(n)$ and denote the $S(n)$-orbit of any
 $\tau\in \bT(n)^0$ by $[\tau]$, then
the action (\ref{A34}) takes the form
\be
\mu: (\zeta_0, [\tau]) \mapsto
(e^{-\ri \frac{2\pi}{n}} \zeta_0, [e^{\ri \frac{2\pi}{n}} \tau]).
\label{A36}\ee
In this picture the projection $K(n)  \to Q(n)$ associated with (\ref{A33}) can be written simply as
\be
 U(1) \times SQ(n)\ni (\zeta_0, [\tau]) \mapsto [\zeta_0 \tau]\in Q(n).
\label{A37}\ee
The cotangent lift of this projection yields the map
\be
\psi_1^\mI: T^*K(n)\simeq T^*U(1) \times T^*SQ(n)  \to T^* Q(n),
\label{A38}\ee
 whereby $(\psi_1^\mathrm{I})^* (\Omega_{T^*Q(n)}) = \Omega_{T^* K(n)}$.
 The notation $\psi_1^\mI$ conforms with diagram (\ref{D}).
Since $T^*K(n)$ is a $\bZ_n$ symplectic covering of $T^*Q(n)$,
one can study the dynamics on $T^*Q(n)$ either by working on $T^*K(n)$ and then projecting to $T^* Q(n)$, or
by directly working  on the non-trivial manifold $Q(n)$.
In the latter approach one may use the coordinates introduced below.

\subsection{Convenient coordinates on $Q(n)$}

We here construct a cover of $Q(n)$  by two contractible coordinate charts.
For this purpose, we regard  the elements of $Q(n)$ as $S(n)$-orbits in $\bT(n)^0$ (\ref{A33}) and also use of the submersion
$\det: Q(n) \to  U(1)$ defined by
\be
\det ([X]):= \det(X),
\quad
\forall X\in \bT(n)^0.
\label{A39}\ee
For any $z\in U(1)$, the inverse image $\det^{-1}(z) \subset Q(n)$
consists of the $S(n)$-orbits
$[X]$ for which  $\det([X])=z$.
If $z^{1/n}$ is an $n$-th root of $z\in U(1)$, then any
$[X]\in Q(n)$ `over $z$' is of the form
$[X] = [z^{1/n} Y]$
for uniquely determined
$[Y] \in SQ(n)$.
In this way, a choice of $z^{1/n}$ gives rise to a diffeomorphism
between  $\det^{-1}(z)\subset Q(n)$ and $SQ(n)$.
As a result, we see that
\be
(Q(n), U(1), SQ(n), \det)
\label{A40}\ee
is a fiber bundle over base $U(1)$ and fiber type
given by $SQ(n)$.

Let us cover the unit circle  $U(1)$ with two
coordinate charts
\bea
&&
\cU :=  \{ e^{\ri \phi}\,\vert\,
-\epsilon < \phi < \pi + \epsilon\} \simeq (-\epsilon,\pi + \epsilon),\nonumber\\
&&
\cU^\prime :=  \{ e^{\ri \phi'}\,\vert\,
-\pi-\epsilon < \phi' <  \epsilon\}\simeq (-\pi-\epsilon, \epsilon),
\label{A41}\eea
using some small $\epsilon>0$.
We can trivialize both  $Q(n)\vert_{\cU} :={\det}^{-1}(\cU)$
and $Q(n)\vert_{\cU'}:={\det}^{-1}(\cU')$.
Working
over $\cU$ we
write any $[X]$ with $\det ([X])=e^{\ri \phi}$ in the form
\be
[X] = [ e^{\ri \phi/n} Y],
\quad
[Y]\in SQ(n),
\label{A42}\ee
and  define the  trivialization
$\chi:  Q(n)\vert_{\cU} \to \cU \times SQ(n)$
by
\be
\chi: [X]\mapsto (\det([X]), [Y]).
\label{A43}\ee
Similarly, over  $\cU'$ we can
write any $[X]$ with $\det ([X])=e^{\ri \phi'}$ in the form
\be
[X] = [ e^{\ri \phi'/n} Y'],
\quad
[Y']\in SQ(n),
\label{A44}\ee
and define the trivialization
$\chi':  Q(n)\vert_{\cU'} \to \cU' \times SQ(n)$
by
\be
\chi': [X]\mapsto (\det([X]), [Y']).
\label{A45}\ee
The intersection is the disjoint union $Q(n)\vert_{\cU \cap \cU'}  = Q(n)\vert_{\cV_+}  \sqcup Q(n)\vert_{\cV_-}$,
where $\cV_{\pm}$ are the connected components of $\cU\cap \cU'$:
\be
\cV_+ =\{ e^{\ri \phi}\,\vert\, -\epsilon < \phi < \epsilon\},
\qquad
\cV_-= \{ e^{\ri \phi}\,\vert\, \pi -\epsilon <\phi < \pi + \epsilon\}.
\label{A46}\ee
We find that on the overlap the two trivializations are related as follows:
\be
[Y'] = [Y]
\quad
\hbox{if}\quad
\det ([X]) \in\cV_+
\quad\hbox{and}\quad
[Y'] = [e^{\ri 2\pi/n} Y] \quad
\hbox{if}\quad \det ([X]) \in \cV_-.
\label{A47}\ee
To be more explicit, we identify the fiber $SQ(n)$ with the open simplex
$\operatorname{Simp}_{n-1}$, and then obtain the two coordinate charts
\be
Q(n)\vert_{\cU}\simeq \cU \times \operatorname{Simp}_{n-1} \simeq \{(\phi, \delta)\}
\quad\hbox{and}\quad Q(n)\vert_{\cU'}\simeq \cU' \times \operatorname{Simp}_{n-1} \simeq \{ (\phi', \delta')\}.
\label{A48}\ee
The two systems of coordinates coincide on
$Q(n)\vert_{\cV_+}$, and they are related by
\be
(\phi', \delta'_1,\ldots, \delta'_{n-2}, \delta'_{n-1} )=
(\phi - 2\pi, \delta_2, \ldots, \delta_{n-1}, 2\pi - \sum_{k=1}^{n-1}\delta_k)
\quad
\hbox{on}\quad Q(n)\vert_{\cV_-}.
\label{A49}\ee
Finally,  $T^*Q(n)$ is covered by the two charts
$T^* Q(n)\vert_{\cU}\simeq T^*(\cU \times \operatorname{Simp}_{n-1})$ and
$T^* Q(n)\vert_{\cU'}\simeq T^*(\cU' \times \operatorname{Simp}_{n-1})$ with respective canonical coordinates
$(\phi, p_\phi, \delta, \gamma)$ and $(\phi', p_{\phi'}, \delta', \gamma')$, for which
\be
\Omega_{T^* Q(n)\vert_{\cU}}= d p_\phi \wedge d \phi + \sum_{j=1}^{n-1} d \gamma_j \wedge d \delta_j,
\quad
\Omega_{T^* Q(n)\vert_{\cU'}}= d p_{\phi'} \wedge d \phi' + \sum_{j=1}^{n-1} d \gamma'_j \wedge d
\delta'_j.
\label{A50}\ee
These two systems of canonical coordinates coincide on $T^* Q(n)\vert_{\cV_+}$ and
on $T^*Q(n)\vert_{\cV_-}$ (\ref{A46}) their relation is
\be
(\phi', p_{\phi'}, \delta', \gamma') = (\phi- 2\pi, p_\phi, \delta_2,
\ldots, \delta_{n-1},
2\pi - \sum_{j=1}^{n-1} \delta_j, \gamma_2 - \gamma_1, \ldots, \gamma_{n-1} - \gamma_1, -\gamma_1).
\label{A51}\ee

We finish with two remarks.
First, observe from (\ref{A47}) that bundle $Q(n)$ (\ref{A40}) can be viewed as  an associated bundle
to a principal $\bZ_n$-bundle over $U(1)$, with sewing
function equal to $1 \in \bZ_n$ on $\cV_+$ and equal to the
constant $e^{\ri 2\pi/n}\in \bZ_n$ over $\cV_-$.
This principal $\bZ_n$-bundle is topologically non-trivial.
Second, one may check by computing the Jacobian of the coordinate change (\ref{A49})
that $Q(n)$ is orientable if and only if $n$ is odd, which was shown previously by a different argument in \cite{FK3}.

\bigskip
\medskip
\noindent{\bf Acknowledgements.}
L.F.~thanks J.~Balog, C.~Klim\v c\'\i k and S.N.M.~Ruijsenaars for discussions.
This work was supported in part
by the Hungarian
Scientific Research Fund (OTKA) under the grant K 77400.

\end{document}